\documentclass[aps,prb,twocolumn,notitlepage,superscriptaddress]{revtex4-2}

\usepackage{amsmath}
\usepackage{amssymb}
\usepackage{amsfonts}
\usepackage{graphicx}
\usepackage{bm}
\usepackage{epstopdf}
\usepackage[colorlinks=true]{hyperref}
\usepackage{tensor}

\begin{document}
\title{Magnon dynamics in a Skyrmion-textured domain wall of antiferromagnets}

\date{\today}

\author{Seungho Lee}
\affiliation{Department of Physics, Korea Advanced Institute of Science and Technology, Daejeon 34141, Republic of Korea}
\author{Kouki Nakata}
\affiliation{Advanced Science Research Center, Japan Atomic Energy Agency, Tokai, Ibaraki 319-1195, Japan}
\author{Oleg Tchernyshyov}
\affiliation{Department of Physics and Astronomy, Johns Hopkins University, Baltimore, Maryland 21218, USA}
\author{Se Kwon Kim}
\affiliation{Department of Physics, Korea Advanced Institute of Science and Technology, Daejeon 34141, Republic of Korea}

\begin{abstract}
We theoretically investigate the interaction between magnons and a Skyrmion-textured domain wall in a two-dimensional antiferromagnet and elucidate the resultant properties of magnon transport. Using supersymmetric quantum mechanics, we solve the scattering problem of magnons on top of the domain wall and obtain the exact solutions of propagating and bound magnon modes. Then, we find their properties of reflection and refraction in the Skyrmion-textured domain wall, where magnons experience an emergent magnetic field due to its non-trivial spin texture-induced effective gauge field. Based on the obtained scattering properties of magnons and the domain wall, we show that the thermal transport decreases as the domain wall's chirality increases. Our results suggest that the thermal transport of an antiferromagnet is tunable by modulating the Skyrmion charge density of the domain wall, which might be useful for realizing electrically tunable spin caloritronic devices.
\end{abstract}

\maketitle

\section{Introduction}

\begin{figure}[t]
\includegraphics[width=1\columnwidth]{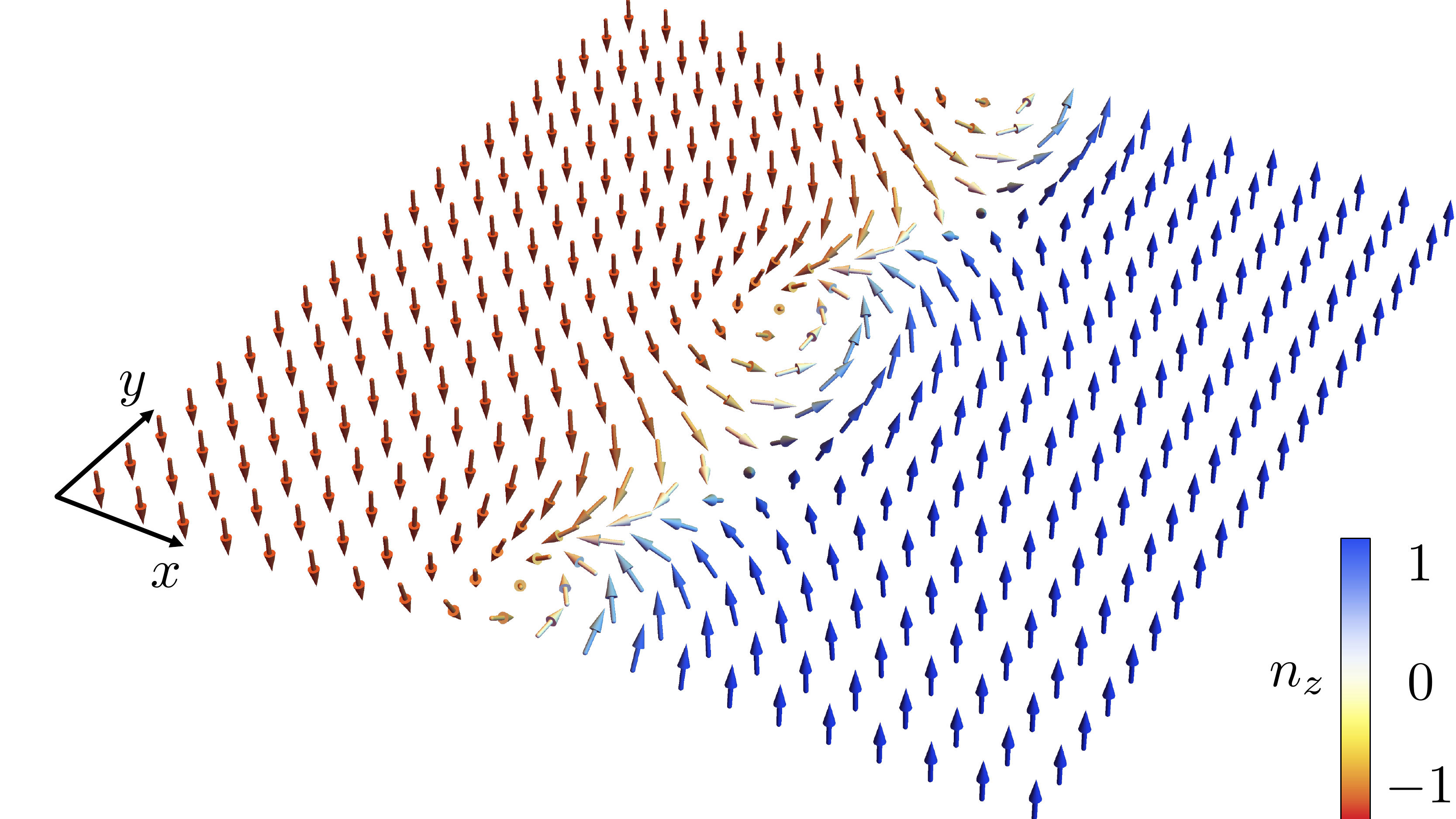}
\caption{A Skyrmion-textured domain wall. The arrows represent the N\'eel order parameter $\mathbf{n}$. The color represents the $z$-component of the N\'eel order parameter.}
\label{skydw}
\end{figure}

Antiferromagnets are arising platforms for spintronic applications due to their exceptional features~\cite{Baltz2018,Jungwirth2016}. An unit cell of an antiferromagnet is composed of two sublattice magnets aligned antiparallelly, which removes the net magnetic moment and thereby makes antiferromagnets robust against external magnetic stimuli. Moreover, the absence of the net magnetic moment allows us to ignore the effects of stray fields which have been hampering the densification of ferromagnetic devices. Compared to ferromagnets whose dynamics are on the gigahertz scale, the inherent dynamics of antiferromagnets are in the range of terahertz that supports the antiferromagnet a candidate of an ultrafast computation platform~\cite{Cheng2015,Khoshlahni2019}.

Elementary magnetic excitations of an antiferromagnetic system exhibit left- and right-circular spin waves~\cite{Gongyo2016,Rezende2019,Hongo2021b,Wang2021}. A quantum of the spin wave is a spin-1 boson and is called a magnon~\footnote{In our work, we use two terms “spin wave” and “magnon” interchangeably with emphasis on wave-like and particle-like properties of magnetic excitations, respectively.}. The magnon current can be free from the electron current which is the classic information carrier of electronics but inevitably implies the Ohmic loss. Therefore magnons are considered as promising candidates of information carriers for low-energy-based devices~\cite{Chumak2015}. Not only fundamental interest for excitations of ordered magnetic system but also potential technological usefulness derives research branches such as magnonics~\cite{Barman2021} and spin caloritronics~\cite{Bauer2012a}.

Topological solitons are topologically classified stable solutions of field theory~\cite{Manton2004a,Vachaspati2006,Weinberg2012,Shnir2018,Nitta2022}. They are found and studied in research areas such as nuclear physics~\cite{Vento2017,Park2019a}, soft matter physics~\cite{Ackerman2015,Ackerman2017a,Afghah2017a,Long2021a,Long2021,Schimming2022,Schimming2020a}, optics~\cite{Parmee2022,Poy2022}, and condensed matter physics~\cite{Lin2016,Psaroudaki2017,Diaz2019,Diaz2020a,Psaroudaki2021,Hayami2021b,Psaroudaki2022,Naya2022,Hayami2022e,Back2020}. One of the mostly studied topological solitons is the Skyrmion~\cite{Skyrme1961a,Skyrme1962}. The Skyrmion was suggested in early 1960's and has been considered as a key to describe baryonic matters in nuclear physics~\cite{Witten1983,Witten1983a}. The originally suggested Skyrmion which resides in the three-dimensional space is defined by the homotopy group $\pi_3(S^3)$. A magnetic Skyrmion is the two-dimensional version of the original Skyrmion, called the baby Skyrmion in high energy community, and defined by $\pi_{2}(S^2)$. In condensed matter community, magnetic Skyrmions are simply called Skyrmions. Similarly Skyrmions are generalized to $N$-dimension space and defined by $\pi_N(S^N)$~\cite{Nitta2013}. From now on we use the term ``Skyrmion'' to refer to the magnetic Skyrmion.

Two-dimensional antiferromagnets support topological solitons such as Skyrmions and domain walls~\cite{Smejkal2018,Kravchuk2019,Hongo2020,Gobel2021}. Skyrmions are point-likely localized in two-dimension and domain walls are point-likely localized in one-dimension~\cite{Nitta2013a,Nitta2022}. These localities have an advantage in information technology~\cite{Xia2017,Hoffmann2021a,Parkin2008}. Furthermore, toplogical solitons can be used as controllers of magnon currents via the interaction of magnons and topological solitons~\cite{Yu2021b}. In particular, a magnon on non-trivial spin textures feels the gauge field so that experiences the effective Lorentz force which generates the transverse dynamics of the magnon~\cite{Schutte2014a,Schroeter2015,Guslienko2016,Tatara2019a}. The non-trivial textures 
frequently become the essence of the transverse transports and also determine the non-trivial properties of the longitudinal transport~\cite{Cheng2012,vanHoogdalem2013,Li2021b}.

In this paper, we investigate the interaction of magnons and a Skyrmion-textured domain wall, which is a topological soliton having properties of the Skyrmion and the domain wall~\cite{Lee2021, Lee2022}, in two-dimensional antiferromagnets. The Skyrimon-textured domain wall shows a chiral texture along the domain wall represented by the topological charge density and determined by boundary conditions that is illustrated in Fig.~\ref{skydw}.  Domain walls with no spin texture along with them are known to be transparent to magnons~\cite{Yan2011a,Kim2014}. However, the Skyrmion-textured domain wall is no longer transparent, since the chiral texture makes a reflective potential barrier for the magnon~\cite{Lee2022}. To obtain the reflection probability, we use supersymmetric quantum mechanics (SUSY QM)~\cite{Sukumar1985,Cooper1995}. We also obtain the exact solutions of magnon-bound modes in the vicinity of the domain wall. Based on the Lagrangian formalism, we derive the gauge field for the magnon and interpret the magnon refraction as a deflection of the magnon trajectory due to the emergent magnetic field. 

 Reflection and refraction phenomena affect the thermal transport of the sample, and thus the thermal transport is vividly chirality-dependent. Generically, the thermal transport is determined by the material parameters of a given sample which are not easy to tune rapidly. Here using the chiral texture, we propse potentially useful means to tune the thermal transport. To elucidate the tunable thermal transport, we use domain walls in an easy-axis antiferromagnet. The easy-axis anisotropy breaks the spin O(3) symmetry down to U(1) and the domain wall spontaneously breaks this U(1) symmetry. In a two-dimensional system, a domain wall is a one-dimensional object and spin textures can spatially vary along the domain wall when chirality is injected~\cite{Go2022}. The chirality injection is tunable by the spin Hall effect of a metal contact at the boundary of the domain wall. This tunability of the spin chirality manifests the tunable thermal transport.

This paper is organized as follows. We begin in Sec.~\ref{sec2} by formulating the field theory for two-dimensional antiferromagnets and introduce the Skyrmion-textured domain wall. Section~\ref{sec3} is devoted to understand the interaction of magnons and the Skyrmion-textured domain wall via SUSY QM and the emergent electromagnetism. In Sec.~\ref{sec4}, we show the chirality dependence of the thermal transport. In Sec.~\ref{sec5}, we summarize and conclude our work. In Appendix~\ref{eompb}, we derive equations of motion for the antiferromagnet via the Poisson bracket. Appendix~\ref{applag} provides detailed calculations of the gauged sigma model~\cite{Tchrakian1995,Schroers2019,Speight2020}. From the gauged sigma model approach, we can naturally see how the gauge field of the magnetic order generates the gauge field of the magnetic excitation.

\section{Skyrmion-textured domain walls in a 2D antiferromagnet}\label{sec2}
In this section, using the continuum field theory of Lagrange-Hamilton formalism, we formulate a two-dimensional antiferromagnet and introduce the Skyrmion-textured domain wall.

\subsection{General formalism}

We consider a two-dimensional collinear antiferromagnet with easy-axis anisotropy, where the magnetizations of two constituent sublattices are antiferromagnetically coupled. The state of the antiferromagnet is represented by the N\'eel order parameter $\mathbf{n}(\mathbf{x}, t)$ which is a three-dimensional unit vector in the direction of the staggered magnetization, i.e., the difference of the magnetization between the two sublattices. The Lagrangian of the system is a functional of the order parameter $\mathbf{n}$, whose density is given by~\cite{Ivanov1983,Haldane1983,Haldane1983a}
\begin{equation}\label{Lag}
	\mathcal{L}=\frac{1}{2} \left[ \rho \left\vert\dot{\mathbf{n}}\right\vert^2 -A\sum_i\left\vert\partial_i\mathbf{n}\right\vert^2-K \left[1-(\mathbf{n}\cdot\hat{z})^2\right]\right] \, ,
\end{equation}where $\rho, A$, and $K$ are the inertia of the staggered magnetization, the exchange coefficient and the anisotropy coefficient, respectively. For the subsequent theoretical discussion, it is convenient to use the natural unit of length, time, and energy by setting $\rho=A=K=1$, in which the Lagrangian density is given by
\begin{equation}
	\mathcal{L}=\frac{1}{2} \left[  \left\vert\dot{\mathbf{n}}\right\vert^2 -\sum_i\left\vert\partial_i\mathbf{n}\right\vert^2-\left[1-(\mathbf{n}\cdot\hat{z})^2\right]\right] \, .
\end{equation}Since the order parameter $\mathbf{n}$ has unit length, it can be expressed by two fields $\theta$ and $\phi$ by 
\begin{eqnarray}
	\mathbf{n}(\mathbf{x},t)&=&(\sin{\theta}\cos{\phi},\sin{\theta}\sin{\phi},\cos{\theta})\,,\label{neel}
	\\
	\theta&=&\theta(\mathbf{x},t)\,,\,\phi=\phi(\mathbf{x},t) \, .
\end{eqnarray}
Since $|\dot{\mathbf{n}}|^2-\sum_i |\partial_i \mathbf{n}|^2=\partial_{\mu} \mathbf{n}\cdot \partial^{\mu} \mathbf{n}=\partial_{\mu}\theta\partial^{\mu}\theta+\sin^2\theta(\partial_{\mu}\phi\partial^{\mu}\phi)$ from Eq.~\eqref{neel},
we obtain the Lagrangian in terms of two fields $\theta$ and $\phi$ 
\begin{equation}
	\mathcal{L}=\frac{1}{2}\left[\partial_{\mu}\theta\partial^{\mu}\theta+\sin^2\theta(\partial_{\mu}\phi\partial^{\mu}\phi)-\sin^2\theta\right] \, .
\end{equation}
Here, we use the Einstein's summation convention and the metric signature is $[+,-,-]$ (2+1 spacetime metric). The following coupled equations of motion for the fields $\theta$ and $\phi$ are obtained from the Euler-Lagrange equations:
\begin{eqnarray}
	\partial_{\mu}\partial^{\mu}\theta-
	\sin\theta\cos\theta\left(\partial_{\mu}\phi\partial^{\mu}\phi-1\right)=0 \, , \label{ELeq}
	\\
	\partial_{\mu}\left(\sin^2\theta\partial^{\mu}\phi\right)=0 \, . \label{spincurrent}
\end{eqnarray}
Equation~\eqref{spincurrent} can be cast into $\partial_\mu j^\mu = 0$, which represents the conservation of spin rooted in the U(1) spin-rotational symmetry of the Lagrangian about the $z$-axis with
\begin{equation}\label{jmu}
	j^\mu=\sin^2\theta \partial^\mu \phi\,.
\end{equation}
The spin density and the spin current density are given by $j^0$ and $\mathbf{j} = (j^x, j^y)$, respectively. Since the Lagrangian has no explicit dependence on space-time, the energy-momentum tensor in terms of $\theta$ and $\phi$, which is given by
\begin{equation}
	T^{\mu\nu}=\partial^\mu\theta\partial^\nu\theta+\sin^2\theta \partial^\mu\phi\partial^\nu\phi-g^{\mu\nu}\mathcal{L}\,,
\end{equation}
is also a conserved current satisfying the continuity equation $\partial_\mu T^{\mu\nu}=0$.

The system can also be studied within the Hamiltonian formalism. Performing the Legendre transformation to the Lagrangian yields the following Hamiltonian density:
\begin{equation}
	\mathcal{H}=\frac{1}{2}\bigg[\pi_\theta^2+\left\vert\bm\nabla \theta\right\vert^2+\frac{\pi_\phi^2}{\sin^2\theta}+\sin^2\theta\left(\left\vert\bm\nabla \phi\right\vert^2 +1\right)\bigg]\,,\label{H}
\end{equation}
where $\pi_\theta\equiv \partial\mathcal{L}/\partial\dot\theta= \dot\theta$ and $\pi_\phi\equiv\partial \mathcal{L}/\partial\dot\phi=\sin^2\theta\dot\phi$ are conjugate momenta of the fields $\theta$ and $\phi$, respectively~\cite{Hasselmann2006}. In the Hamiltonian formalism, the time evolution of fields or momenta is determined by their Poisson brackets with the system's Hamiltonian, which leads us to the same Eq.~\eqref{ELeq} and Eq.~\eqref{spincurrent} as in the Lagrangian formalism. See Appendix~\ref{eompb} for the derivation based on the Hamiltonian formalism.

\subsection{Skyrmion-textured domain walls}

To obtain a static domain-wall solution, we set $\dot{\theta} = 0$ and $\dot{\phi} = 0$ in the equations of motion:
\begin{eqnarray}
	\nabla^2\theta-\sin\theta\cos\theta\left(\boldsymbol\nabla\phi\cdot\boldsymbol\nabla\phi+1\right)=0 \, ,
	\\
	\boldsymbol\nabla\cdot\left(\sin^2\theta\boldsymbol{\nabla}\phi\right)=0 \, .
\end{eqnarray}
For a domain wall, we consider the following ansatz by employing separation of variables:
\begin{eqnarray}
	\theta(x,y)=\theta(x)\,,\, \phi(x,y)&=&\phi(y) \, ,
	\\
	\mathbf{n}(x=\pm\infty,y)&=&\pm\hat{z} \, .\label{bc}
\end{eqnarray}
The equations are solved by the following solution:
\begin{equation}\label{dwsol}
	\cos\theta=\tanh\left(\sqrt{1+k_0^2}x\right) \,  , \quad \phi=k_0y \, .
\end{equation}
Here, $k_0$ is a real number which characterizes chirality of the domain-wall texture. The domain wall interpolates two discrete vacua $\mathbf{n}(x,y)=\pm \hat{z}$. In a two-dimensional system, a domain wall is a one-dimensional object which can exhibit nontrivial textures along with it~\cite{Kobayashi2013a}. With the considered solution, domain-wall angle $\phi$ varies uniformly along the domain wall (i.e., along the $y$-axis, see Fig.~\ref{skydw}), which gives rise to the finite Skyrmion charge density given by
\begin{eqnarray}
	\rho_\text{Sky}(x,y)&\equiv&\frac{1}{4\pi}\mathbf{n}\cdot\left(\partial_x\mathbf{n}\times\partial_y\mathbf{n}\right)
	\\
	&=&-\frac{k_0}{4\pi}\left(1+k_0^2\right)\text{sech}^2\left(\sqrt{1+k_0^2 }x\right)\,.
\end{eqnarray}
Note that the domain wall carries the Skyrmion charge density, which is localized in the vicinity of the domain wall. Along the $y$-axis, the density is uniform and the linear Skyrmion charge density (per unit length in the $y$-direction) is given by
\begin{equation}
	\int dx \rho_\text{Sky}=-\frac{k_0}{2\pi}\,.
\end{equation}
If the periodic boundary condition along the $y$-axis is demanded ($y+L=y$), we would have $\phi(L)=\phi(0)$, which discretizes the allowed values for $k_0$ as  $k_0=2\pi n /L$ for an integer $n$. The integration of the density yield the integer Skyrmion charge $\int dx dy \rho_\text{Sky} = -n $~\cite{Lee2021}.
In the case of $k_0=0$, the domain-wall profile along the $y$-axis is uniform and the corresponding Skyrmion charge is zero.

For the static field, the Hamiltonian density is given by $\mathcal{H}=-\mathcal{L}$ . The explicit expression for the energy density of the domain-wall solution~\eqref{dwsol} is given by
\begin{equation}
	\mathcal{H}=\left(1+k_0^2\right)\text{sech}^2\left(\sqrt{1+k_0^2 }x\right)\,.
\end{equation}
The energy density per unit length in the $y$-axis is given by
\begin{equation}
	E=\int dx \mathcal{H} =2\left(1+k_0^2\right)^\frac{1}{2}\,.
\end{equation}
Note that the nontrivial magnetic textures on the domain wall increases the  energy.

Experimentally, the obtained domain-wall states with the finite Skyrmion charge can be realized by injecting a spin current into the magnet via the spin Hall effects as shown in Ref.~\cite{Kim2017a} for a ferromagnet and in Ref.~\cite{Flebus2018,Flebus2019} for an antiferromagnet.

\section{Spin waves on top of the Skyrmion-textured domain wall}\label{sec3}
In this section, we formulate the dynamics of magnons on top of the Skyrmion-textured domain wall. With the aid of SUSY QM, we find chiral bound modes. Furthermore, we obtain an analytic form of a reflection probability of the magnon and show their refraction in the presence of the skyrmionic texture of the domain wall. This refraction is interpreted from a perspective of the emergent magnetic field for magnons.
\subsection{Spin waves}
To obtain a spin-wave solution on top of the Skyrmionic domain wall, we consider a small fluctuation $\delta \mathbf{n}(\mathbf{x}, t)$ on the static domain-wall background $\mathbf{n}(x)$ [Eq.~(\ref{dwsol})]. The variation of the vector is described by variations of the angles:
\begin{equation}\label{n1n2}
	\delta n_1 = \delta \theta \,, \quad \delta n_2 = \sin\theta \delta \phi \,.
\end{equation}
Linearizing the Euler-Lagrange equations [Eqs.~\eqref{ELeq} and \eqref{spincurrent}] with respect to the variation $(\theta,\phi) \mapsto (\theta +\delta \theta,\phi+\delta \phi)$, we obtain two coupled equation for $\delta n_1 $ and $ \delta n_2$. The two equations are combined into one complex equation
\begin{equation}\label{eom}
	\left[\square+\left(1+k_0^2\right)\left(1-2\sin^2\theta\right)+2qik_0\cos\theta\partial_y\right] \Psi_q=0\,,
\end{equation}
where $\Psi_q \equiv \delta n_1 -q i \delta n_2$, $q=\pm 1$, and $\square \equiv \partial_\mu \partial^\mu$. Here, the complex field $\Psi_q$ represents $q$-polarized spin waves, where $q = 1$ and $q = -1$ are for right-handed and left-handed spin waves, respectively~\cite{Li2020}. Due to the translational symmetries with respect to $(t,y)\mapsto (t+\delta t , y + \delta y) $ of the equation,
\begin{equation}\label{psi}
	\Psi_q(t,x,y)= \psi_q(x) e^{i\left(k_y y - \omega t \right)}
\end{equation}
is supported to be an eigenfunction. Equation of motion for the eigenfunction is given by 
\begin{equation}
\begin{aligned}
	\omega^2 \psi_q = \big[& -\partial_x^2 + k_y^2 + \left(1+k_0^2\right)\left(1-2\sin^2\theta\right) \\ & - 2qk_0k_y\cos\theta\big] \psi_q \, .
\end{aligned}
\end{equation}
We interpret the equation as the ``Schr\"odinger'' equation $\omega^2 \psi_q = H \psi_q$ with the eigenvalue $\omega^2$ and the ``Hamiltonian'' $H=-\partial_x^2 + V$, which consists of the kinetic energy $-\partial_x^2$ and the potential energy given by
\begin{equation}\label{potential}
\begin{aligned}
	V=&k_y^2 +\left(1+k_0^2\right)\left[1-2\text{sech}^2\left(\sqrt{1+k_0^2}x\right)\right]
	\\
	&-2qk_0k_y\tanh\left(\sqrt{1+k_0^2}x\right)\,.
\end{aligned}
\end{equation}
Equation~\eqref{eom} with the potential~\eqref{potential} is one of our main results. Here, the potential energy $V$ is known as the Rosen-Morse potential~\cite{Rosen1932}. Note that, for the case of the non-textured domain wall $k_0=0$, the potential becomes the P\"oschl-Teller potential~\cite{Poeschl1933} which is known as a reflectionless potential~\cite{Lekner2007}. In general, with nonzero $k_0$, spin waves are reflected; the derivation of the reflection probability is presented below in Sec.~\ref{secrefl}.

\subsection{SUSY QM}
\begin{figure}[t]
\includegraphics[width=1\columnwidth]{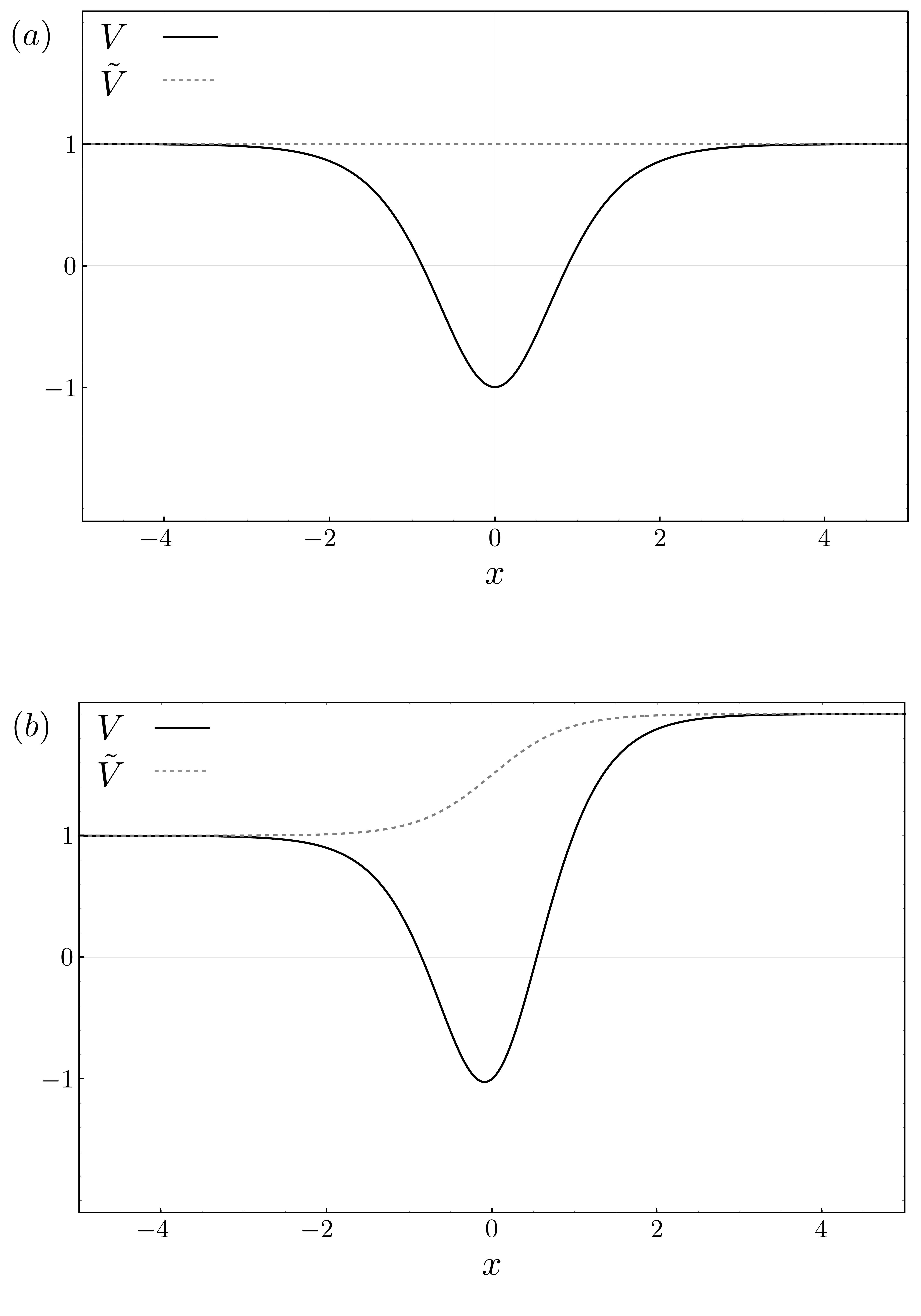}
\caption{ The solid line represents the original potential~\eqref{potential} and the dashed line represents the SUSY partner~\eqref{partpot}. Here, we consider the case of left-polarized magnons coming from the left $x<0$. The chirality of the Skyrmion-textured domain wall are $(a)$ $k_0=0$ and $(b)$ $k_0=0.5$, where $k_y=k_0$.}
\label{fig2}
\end{figure}
In this section, we solve the equation of motion for the spin wave with the aid of SUSY QM. To apply the SUSY QM, we define the annihilation and creation operators
\begin{eqnarray}
	a&=&\partial_x +\sqrt{1+k_0^2} \tanh\left(\sqrt{1+k_0^2} x \right) +\beta\,,
	\\
	a^\dagger&=&-\partial_x +\sqrt{1+k_0^2} \tanh\left(\sqrt{1+k_0^2} x \right) +\beta\,,
\end{eqnarray}
where $\beta=-q k_y k_0 / \sqrt{1+k_0^2}$. Commutation relation of the operators is given by
\begin{equation}\label{commutator}
	[a,a^\dagger]=2\left(1+k_0^2\right)\text{sech}\left(\sqrt{1+k_0^2}x\right)\,.
\end{equation}
With these operators, the Hamiltonian is written as $H=a^\dagger a -\beta^2+ k_y^2$ and the SUSY partner Hamiltonian is defined as $\tilde{H}=aa^\dagger -\beta^2 +k_y^2$. The SUSY partner of the potential is defined by the relation $\tilde{H}=-\partial_x^2 +\tilde {V}$. Since, by the definition, $\tilde{H}=H+[a,a^\dagger]$, the commutator~\eqref{commutator} eliminates the $\text{sech}$ term in the potential~\eqref{potential}. Thus the induced partner potential is simpler than the original potential, that is
\begin{equation}\label{partpot}
	\tilde{V}=k_y^2 +1+k_0^2 -2qk_0k_y\tanh\left(\sqrt{1+k_0^2}x\right)\,.
\end{equation}
Figure~\ref{fig2} provides comparison of the original potential $V$ and the partner potential $\tilde{V}$. Note that the partner potential for the case of $k_0=0$ is a constant potential.
Due to the algebraic relation of $H$ and $\tilde{H}$, eigenfunctions for both Hamiltonians satisfy the following relations:
\begin{eqnarray}
	\tilde{H}a\psi_q&=&a H \psi_q = \omega^2  a \psi_q\,,
	\\
	Ha^\dagger\tilde{\psi}_q&=&a^\dagger \tilde{H}\tilde{\psi}_q=\tilde{\omega}^2 a^\dagger \tilde{\psi}_q\,,
\end{eqnarray}
where $\tilde{H}\tilde{\psi}_q=\tilde{\omega}^2\tilde{\psi}_q$. It means that $a\psi_q(a^\dagger\tilde{\psi}_q)$ is an eignefunction of $\tilde{H}(H)$. We will solve the problem in the partner system which is easier than the original system and obtain the solution for the original problem by applying the ladder operator,  $\psi_q = a^\dagger \tilde{\psi}_q$.

\begin{figure}[t]
\includegraphics[width=1\columnwidth]{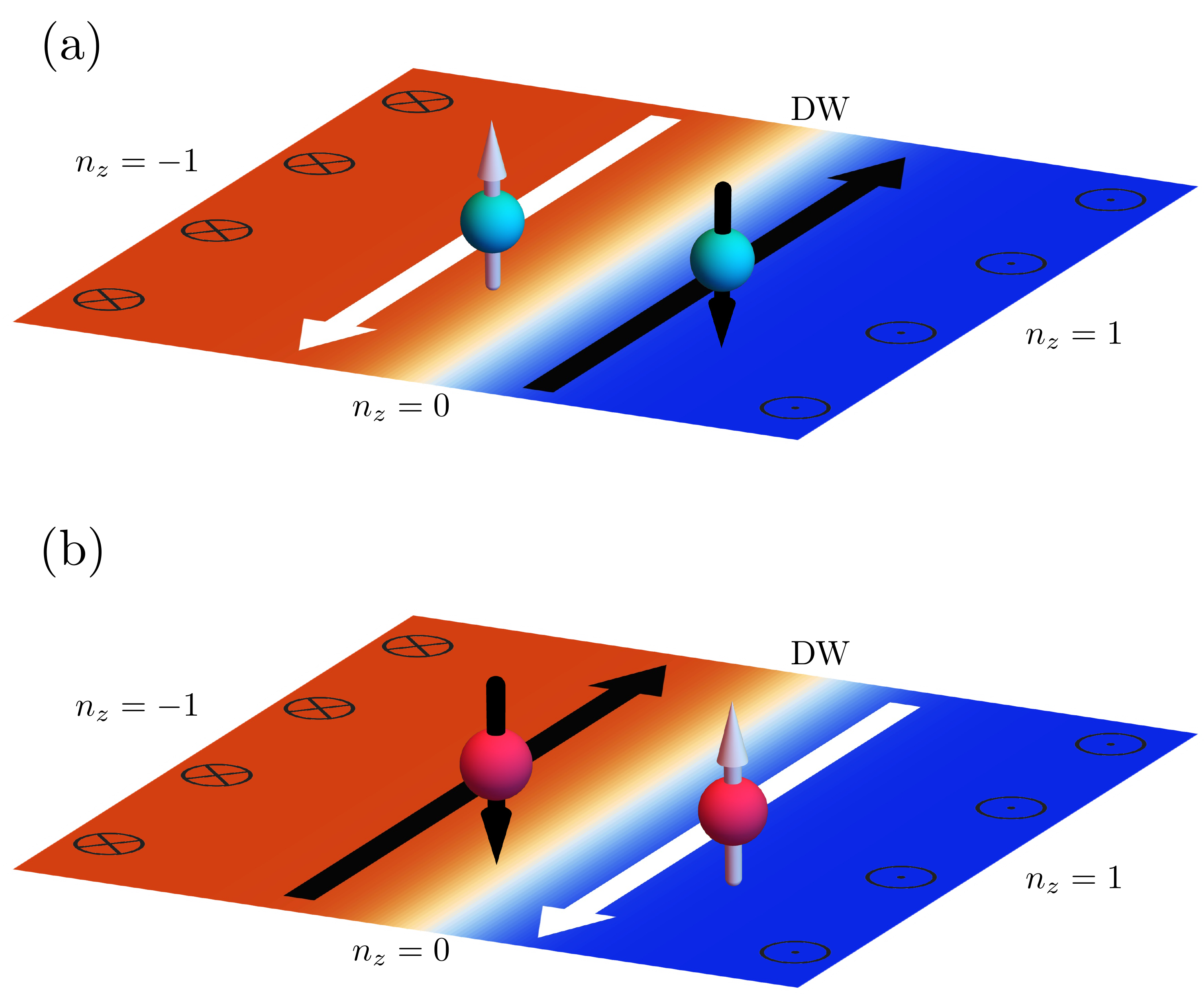}
\caption{ Trajectories of (a) left-polarized bound magnons and (b) right-polarized bound magnons. Here $k_0$ is assumed to be positive. The arrows piercing the balls represent the spins of the magnons. The arrows on the plane indicate the direction of the spin current due to the bound magnons.}
\label{fig3}
\end{figure}
\subsection{Domain-wall bound spin waves}
Before going to elucidate propagating modes, we will discuss bound modes. Since the bound modes satisfy relation $a\psi_q=0$, the solution is given by
\begin{equation}\label{boundmode}
	\psi_q(x)=\text{sech}\left(\sqrt{1+k_0^2} x \right)e^{-\beta x}\,,
\end{equation}
whose bound frequency is $\omega^2=-\beta^2+k_y^2$. The solution for bound modes is one of our main results. Note that, in this paper, the bound magnon propagates to $y$-direction whereas is localized in $x$-direction.  Our solution generalizes the well-known bound solution~\cite{Zhang2018c} which corresponds to the case of $\beta=0$. One can define the bound magnon's position as the location where the amplitude of the wavefunction is maximized. By this definition, the position of the bound magnon is given by
\begin{equation}\label{boundposition}
	X=-\frac{q}{\sqrt{1+k_0^2}}\tanh^{-1}\left(\frac{k_yk_0}{1+k_0^2}\right)\,.
\end{equation}
Equation~\eqref{boundposition} tells us that positions of bound modes with different $k_y$ are seperated and left- and right-polarized bound magnon have opposite positions. Figure~\ref{fig3} schematically illustrates the trajectories of left- and right-polarized bound magnons.

\subsection{Propagating spin waves}\label{secrefl}
In the partner system described by $\tilde{H}$, asymptotic behaviors of the propagating wavefunction are
\begin{eqnarray}
	\tilde{\psi}_q(x\to-\infty)&\sim&e^{ik_{-}x}+re^{-ik_{-}x}\,,
	\\
	\tilde{\psi}_q(x\to+\infty)&\sim&te^{ik_{+}x}\,,
\end{eqnarray}
where $k_{+}^2= k_{-}^2-4k_0k_y$ (from the conservation of the energy) and $k_{-}$ is the wavenumber of the incoming spin wave. By acting of $a^\dagger$ on $\tilde{\psi}_q$, asymptotic behaviors of the propagating wavefunction in the original system are given by
\begin{eqnarray}
	a^\dagger\tilde{\psi}_q(x\to-\infty)&\sim&\left[-ik_{-}-2k_0k_y +\beta \right]e^{ik_{-}x}\nonumber
	\\
	&+&r\left[ik_{-}-2k_0k_y +\beta \right]e^{-ik_{-}x}\,,
	\\
	a^\dagger\tilde{\psi}_q(x\to+\infty)&\sim&t\left[-ik_{+}+2k_0k_y +\beta \right]e^{ik_{+}x}\,.
\end{eqnarray}
Reflection probability of the original potential is
\begin{equation}
	\mathfrak{R}=\left\vert\frac{ik_{-}-2k_0k_y +\beta}{-ik_{-}-2k_0k_y +\beta}\right\vert^2\left\vert r \right\vert ^2=\left\vert r \right\vert ^2 \,,
\end{equation}
and it is known to be equivalent to the reflection probability of the partner potential from SUSY QM. The partner system has the hyperbolic-tangent potential which is known solvable. The transmittion probability $\mathfrak{T}=1-\mathfrak{R}$ is given by~\cite{Gadella2017,Boonserm2011,Kim2014}
\begin{equation}
	\mathfrak{T}(k_-,k_y)=\left[1-\frac{\sinh^2\left[\frac{\pi\left(k_{+}-k_{-}\right)}{2\sqrt{1+k_0^2}}\right]}{\sinh^2\left[\frac{\pi\left(k_{+}+k_{-}\right)}{2\sqrt{1+k_0^2}}\right]}\right]\Theta(k_-^2-4k_0k_y),
\end{equation}
where $\Theta$ is the Heaviside step function.
To capture the deflection of a magnon trajectory in the real space, we express the wavefunction in the laboratory frame $\{\hat{x},\hat{y},\hat{z}\}$
\begin{equation}
\begin{aligned}
	\psi_\text{Lab}^\text{in}=&\delta\mathbf{n}\cdot(\hat{x}-i\hat{y})=(-1)\Psi e^{-ik_0y}
	\quad \text{as} \quad z\to-\infty
	\\
	\sim&\,\psi(x)e^{i\left[(k_y-k_0)y-\omega t\right]}\,,
	\\
	\psi_\text{Lab}^\text{out}=&\delta\mathbf{n}\cdot(\hat{x}+i\hat{y})=\Psi e^{ik_0y}
	\quad \text{as} \quad z\to+\infty
	\\
	\sim&\,\psi(x)e^{i\left[(k_y+k_0)y-\omega t\right]}\, ,
\end{aligned}
\end{equation}
where the extra factor $\exp(\pm i k_0 y)$ comes from the fact that $\Psi$ is defined with respect to the local spin frame where the azimuthal angle changes along the $y$ direction in the domain wall. For the concrete discussion, let us first consider left-polarized magnons. Note that $y$-component of the magnon's linear momentum is changed from $k_y-k_0$ to $k_y+k_0$ after passing by the domain wall. In the laboratory frame, wavevectors of incoming, reflected and transmitted magnons with left-handed polarization are given by
\begin{eqnarray}
	\mathbf{k}^\mathrm{L} &=&\left( k_{-},k_y-k_0,0\right)\,,
	\\
	\mathbf{k}_r^\mathrm{L} &=&\left( -k_{-},k_y-k_0,0\right)\,,
	\\
	\mathbf{k}_t^\mathrm{L} &=&\left( k_{+},k_y+k_0,0\right)\,.
\end{eqnarray}
In the laboratory frame, the kinetic energy of magnons are conserved, which ensures the following equalities: $\left\vert\mathbf{k}\right\vert=\left\vert\mathbf{k}_r\right\vert=\left\vert\mathbf{k}_t\right\vert$. Let us consider a magnon incoming from $x<0$ along the $x$-axis. This is the case of $k_y-k_0=0$. This magnon initially has no $y$-component of wavevector. However, after the transmission through the domain wall, the magnon has non-zero $y$-component $2k_0$. For magnons with right-handed polarization, reflection direction is opposite. Wavevectors of incoming, reflected and transmitted right-polarized magnon are
\begin{eqnarray}
	\mathbf{k}^\mathrm{R}&=&\left( k_{-},k_y+k_0,0\right)\,,
	\\
	\mathbf{k}^\mathrm{R}_r&=&\left( -k_{-},k_y+k_0,0\right)\,,
	\\
	\mathbf{k}^\mathrm{R}_t&=&\left( k_{+},k_y-k_0,0\right)\,.
\end{eqnarray}

 Expanding the spin densitiy $j^0(\theta,\phi)$~\eqref{jmu} with respect to $(\theta \to \theta+\delta\theta,\, \phi \to \phi+\delta\phi)$ up to second order, we can obtain spin of $q$-polarized magnons. Since unperturbed fields $\theta$ and $\phi$ are time-independent, the zeroth order of the spin density is zero. Using the definition~\eqref{n1n2}, we expand the spin density
\begin{equation}\label{expand}
\begin{aligned}
	j^0&=\sin^2\theta \partial_t\delta \phi + 2\sin\theta\cos\theta\delta \theta \partial_t \delta \phi
	\\
	&=\sin\theta \partial_t \delta n_2 + 2 \cos\theta \delta n_1 \partial_t \delta n_2\,.
\end{aligned}
\end{equation}
From the following expression of the magnon wavefunction~\eqref{psi}
\begin{equation}
	\Psi_q(t,x,y)=\delta n_1 -qi\delta n_2=\left| \psi_q(x) \right| e^{i(k_y y -\omega t+\eta)}\,, 
\end{equation}
the perturbative fields read as
\begin{equation}\label{cossin}
\begin{aligned}
	\delta n_1 \propto \cos(k_y y -\omega t+\eta)\,,
	\\
	\delta n_2 \propto -q \sin(k_y y -\omega t+\eta)\,,
\end{aligned}
\end{equation}
where $\eta$ is an arbitrary phase. Pluging Eq.~\eqref{cossin} into Eq.~\eqref{expand} and evaluating its the expectation value with respect to time averaging, the first order term vanishes and the remained term yields
\begin{equation}
	\left\langle j^0 \right\rangle_t \propto  q \omega \cos \theta =  q \omega \tanh\left(\sqrt{1+k_0^2}x\right)\,.
\end{equation}
This equation indicates that spin of the left(right)-polarized magnon is changed from $+(-)\hbar$ to $-(+)\hbar$ while the magnon is passing through the domain wall.

\begin{figure}[t]
\includegraphics[width=0.8\columnwidth]{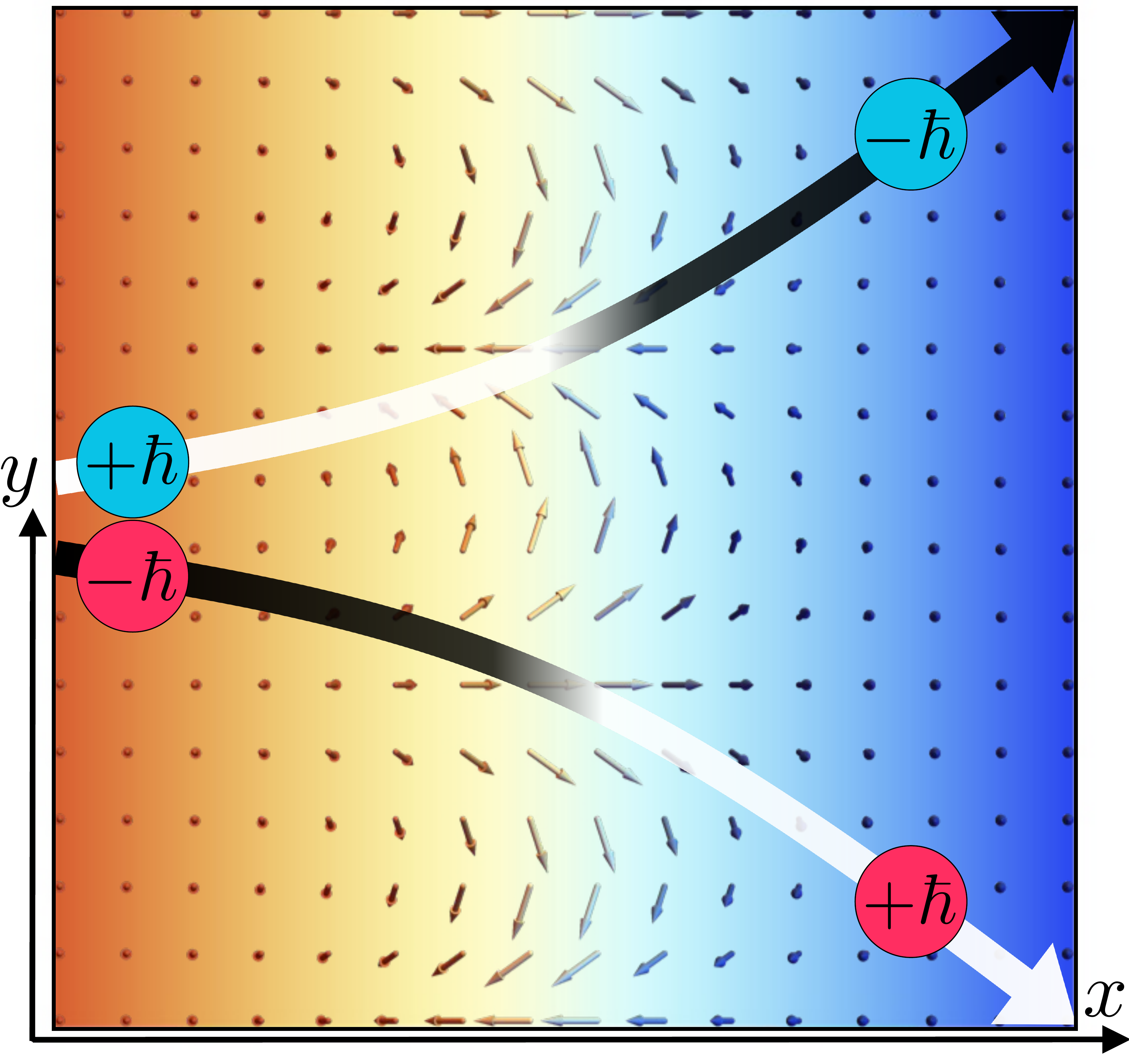}
\caption{Schematic illustration of the motion of distinctly polarized antiferromagnetic magnons moving across the Skyrmion-textured domain wall. Similar to Fig.~\ref{fig3}, the blue(red) ball represents the left(right)-polarized magnon and the color change of the arrow depicts the change of spin angular momentum of the magnon passing through the domain wall [the change from white(black) to black(white) represents the change from $+(-)\hbar$ to $-(+)\hbar$]. Note that, when the same number of two distinctly polarized magnons move in the longitudinal direction, the resultant net transverse magnon current vanishes, but the net transverse spin current is finite.}
\label{fig4}
\end{figure}

\subsection{Perspective of emergent electromagnetism}

Here, we describe an alternative way to understand the deflection of a magnon trajectory by invoking the emergent electromagnetism of spin waves on top of a magnetic texture. From the second-order variation of the Lagrangian of the antiferromagnets, we obtain the Lagrangian of the magnon
\begin{equation}
	\mathcal{L}_\mathrm{SW}=\frac{1}{2}\left(D_\mu \Psi_q\right)^*\left(D^\mu\Psi_q\right)\,,
\end{equation}
where the covariant derivatives and texture-induced gauge field is given by~\cite{Kim2019b,Li2020}
\begin{equation}
	D_{\mu}=\partial_\mu+iqa_\mu\,,\quad a_\mu=-\cos\theta\partial_\mu \phi,.
\end{equation}
Detailed derivation of the Lagrangian can be found in Appendix~\ref{applag}.
From the given domain-wall solution~\eqref{dwsol}, the emergent magnetic field is obtained by~\cite{Kim2017}
\begin{equation}
\begin{aligned}
	b&=-(\partial_1 a_2 -\partial_2 a_1)
	\\
	&=k_0\sqrt{1+k_0^2}\text{sech}^2\left(\sqrt{1+k_0^2}x\right)\,.
\end{aligned}
\end{equation}
The change of the transverse momentum due to the emergent Lorentz force is given by
\begin{equation}
\begin{aligned}
	\Delta p_y=&\int q\left(\mathbf{v}\times\mathbf{b}\right)\cdot \hat{y} dt
	\\
	=&-2qk_0\,,
\end{aligned}
\end{equation}
where $\mathbf{v}$ is the magnon velocity and $\mathbf{b}=b\hat{z}$. This results is consistent to the result obtained by SUSY QM. Note that the change of the transverse momentum depends on the polarization $q$ of the magnon. This indicates that, when the same number of two distinctly polarized magnons pass the Skrymion-textured domain wall in the longitudinal direction (e.g., $x$-direction), there arises a net finite spin current while having a zero net magnon current in the transverse direction (e.g., $y$-direction). The situation is illustrated in Fig.~\ref{fig4}.

\section{Tunable thermal transport}\label{sec4}
In this section, we obtain the chirality-dependent thermal transport. The injected spin current at the boundary which is electronically tunable via the spin Hall effect determines mathematical boundary conditions of the system. Thus we can control the chirality of the texture that is represented by $k_0$~\eqref{dwsol}.
To compute the tunable heat flux, we use the Landauer-B\"uttiker formula~\cite{Shen2020a,Yan2012,Pekola2021}. Heat flux per unit length is given by
\begin{equation}
\begin{aligned}
	\mathcal{J} &= \sum_q \int_0^\infty \frac{dk_x}{2\pi}\int_{-\infty}^{\infty}\frac{dk_y}{2\pi} \mathfrak{T}(k_x,k_y) \hbar
	\\
	&\Bigg[ \omega_{q\mathrm{L}}  n(\omega_{q\mathrm{L}},T_\mathrm{L}) \frac{\partial \omega_{q\mathrm{L}}}{\partial k_x}- \omega_{q\mathrm{R}}  n(\omega_{q\mathrm{R}},T_\mathrm{R}) \frac{\partial \omega_{q\mathrm{R}}}{\partial k_x}\Bigg]\,,
\end{aligned}
\end{equation}
where $\omega_{q\mathrm{L}(\mathrm{R})}$ is the frequecy of a $q$-polarized magnon at the left(right) domain $x<0$($x>0$), $T_{\mathrm{L}(\mathrm{R})}$ is the temperature of the left(right) domain, and $n(\omega,T)$ is the Bose-Einstein distribution.
\begin{figure}[b]
\includegraphics[width=1\columnwidth]{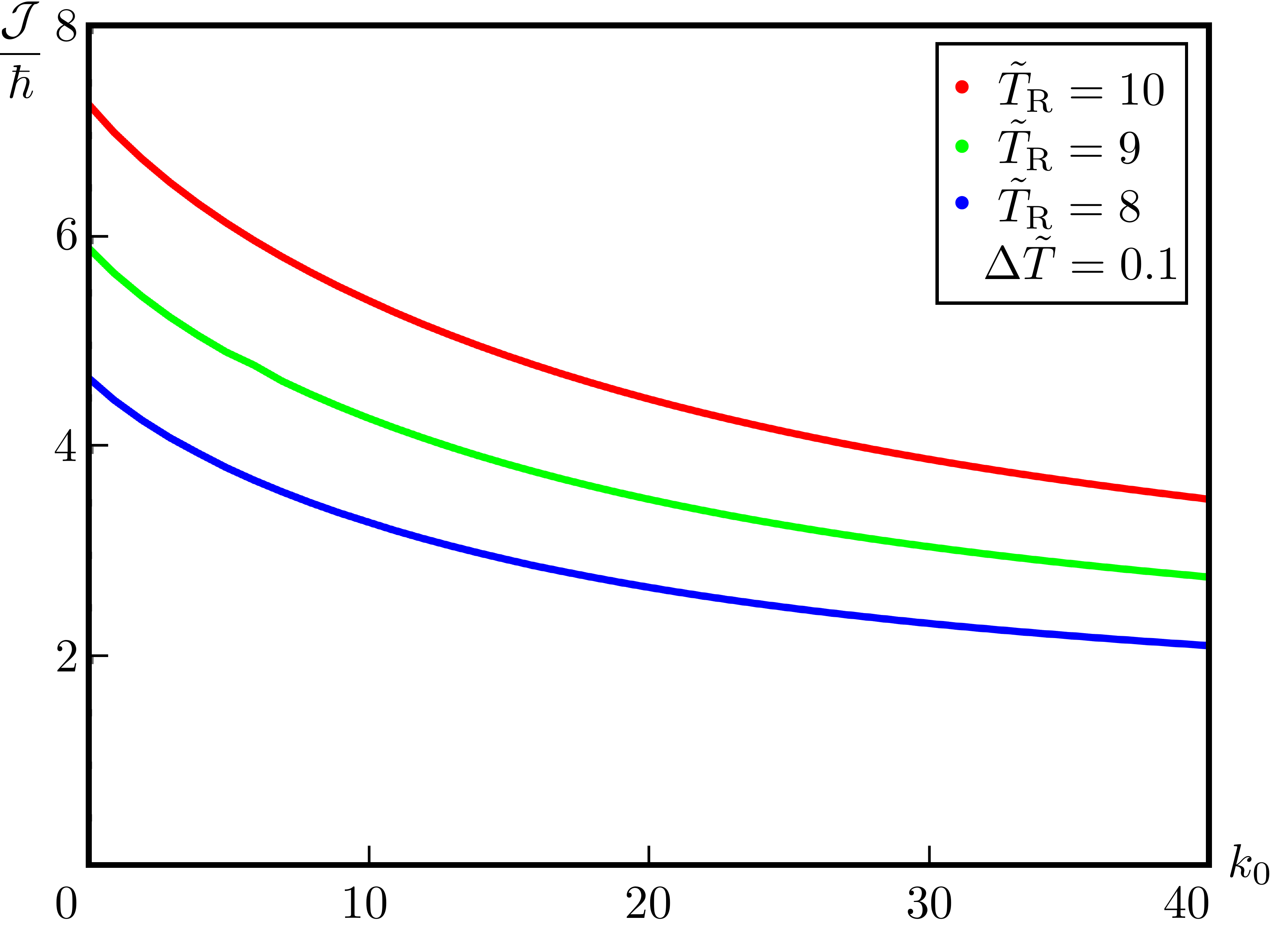}
\caption{Chirality dependence of the heat flux. Here, $\tilde{T}_\mathrm{R}$ denotes the rescaled temperature of the right reservoir and the temperature difference between two reservoirs is fixed as $\Delta \tilde{T}=0.1$, where $\tilde{T}_\mathrm{L}-\tilde{T}_\mathrm{R} \equiv \Delta \tilde{T}$.}
\label{fig5}
\end{figure}
Figure~\ref{fig5} shows chirality dependence of the heat flow. $\tilde{T}_{\mathrm{L}(\mathrm{R})}\equiv  k_\mathrm{B} T_{\mathrm{L}(\mathrm{R})}/ \hbar$ is a rescaled temperature. Note that we use the natural unit $\rho=K=A=1$. Therefore time, length, and energy are measured by $\sqrt{\rho/K}$, $\sqrt{A/K}$, and $A$, respectively. As we expected, chirality of the Skyrmion-textured domain wall disturbs the longitudinal magnon transport.

\section{Conclusion}\label{sec5}

We have formulated a theory of the Skyrmion-textured domain wall in a two-dimensional antiferromagnet. Using the Lagrangian formalism, we have derived equations of motion for magnons on top of the Skyrmion-textured domain wall. Using SUSY QM, we have obtained the exact solutions of magnon bound modes and investigated their scattering properties in the Skyrmion-textured domain wall. With the chiral texture, the domain wall potential is no longer symmetric under the space inversion, and thus the position of the bound magnon is shifted from the domain wall center. Solving the scattering problem of the domain wall potential for magnons, we have shown that magnons are refracted or reflected due to the chiral texture of the domain wall. This refraction can be interpreted as the magnon dynamics under the emergent electromagnetism. Using the gauged sigma model approach, we also have analyzed the Lagrangians of the magnet and the magnon and find how the effective electromagnetism for the magnons emerges in the chiral texture of the background magnet. We have found the total reflection of magnons which are absent of enough longitudinal momentum to overcome the domain wall potential. The total reflection and the chirality-proportional reflection probability which can be tuned electronically reduces the thermal transport.

\begin{acknowledgements}
This work was supported by Brain Pool Plus Program through the National Research Foundation of Korea funded by the Ministry of Science and ICT (NRF-2020H1D3A2A03099291), National Research Foundation of Korea(NRF) grant funded by the Korea government(MSIT) (NRF-2021R1C1C1006273), and National Research Foundation of Korea funded by the Korea Government via the SRC Center for Quantum Coherence in Condensed Matter (NRF-2016R1A5A1008184). K.N. acknowledges support by JSPS KAKENHI Grant Number JP20K14420 and JP22K03519.
\end{acknowledgements}

\newpage

\appendix
\begin{widetext}
\section{Equations of motion from Poisson brackets}\label{eompb}
Poisson brackets of the fields and its momenta are
\begin{eqnarray}
	\left\{\varphi_a(\mathbf{x},t),\,\varphi_b(\mathbf{x}',t)\right\}&=&0\,,
	\\
	\left\{\pi_a(\mathbf{x},t),\,\pi_b(\mathbf{x}',t)\right\}&=&0\,,
	\\
	\left\{\varphi_a(\mathbf{x},t),\,\pi_b(\mathbf{x}',t)\right\}&=&\delta_{ab}\delta(\mathbf{x}-\mathbf{x}')\,.
\end{eqnarray}
Here, indices $a$ and $b$ denote the fields $\theta$ and $\phi$. With these canonical relations, equations of motion for the fields are driven, since time evolution is given by a Poisson bracket with the Hamiltonian. For $\theta$ and $\pi_\theta$, time evolution are
\begin{eqnarray}
	\dot\theta &=& \left\{ \theta,H \right\} 
	= \int d^2x' \left\{\theta(\mathbf{x}),\mathcal{H}\right\}
	= \int d^2x' \left\{\theta(\mathbf{x}), \frac{1}{2} \pi_\theta^2 \right\}
	=\int d^2x' \left\{\theta(\mathbf{x}),\pi_\theta \right\} \pi_\theta
	= \int d^2x' \delta (\mathbf{x}-\mathbf{x}') \pi_\theta(\mathbf{x}')
	\\
	&=&\pi_\theta\,,
	\\
	\dot\pi_\theta&=&\left\{\pi_\theta,H\right\}=\int d^2x' \left\{\pi_\theta(\mathbf{x}),\mathcal{H}\right\}
	\\
	&=&\int d^2x' \Bigg[\left\{\pi_\theta(\mathbf{x}),\frac{1}{2}\bm\nabla\theta\cdot\bm\nabla\theta\right\}
	+\left\{\pi_\theta(\mathbf{x}),\frac{1}{2}\frac{\pi_\phi^2}{\sin^2\theta}\right\}
	+\left\{\pi_\theta(\mathbf{x}),\frac{1}{2}\sin^2\theta\right\}(\bm\nabla \phi\cdot \bm\nabla\phi+1)\Bigg]
	\\
	&=&\int d^2x' \Bigg[\nabla^2\theta (\mathbf{x}')\delta(\mathbf{x}-\mathbf{x}')+\pi_\phi^2\frac{\cos\theta}{\sin^3\theta}\delta(\mathbf{x}-\mathbf{x}')
	-\sin\theta\cos\theta(\bm\nabla \phi\cdot \bm\nabla\phi+1)\delta(\mathbf{x}-\mathbf{x}')\Bigg]
	\\
	&=&\nabla^2\theta+\sin\theta\cos\theta\bigg(\dot\phi^2-\bm\nabla\phi\cdot\bm\nabla\phi-1\bigg)\,.\label{eqpitheta}
\end{eqnarray}
Since $\dot\pi_\phi=\ddot\theta$, Eq.~\eqref{eqpitheta} is consistent to Eq.~\eqref{ELeq}. Similarly equations for $\phi$ and $\pi_\phi$ are
\begin{eqnarray}
	\dot\phi&=&\left\{\phi,H\right\}=\int d^2x' \left\{\phi(\mathbf{x}),\mathcal{H}\right\}
	=\int d^2x' \left\{\phi(\mathbf{x}),\frac{\pi_\phi^2}{2\sin^2\theta}\right\}
	=\int d^2x' \delta(\mathbf{x}-\mathbf{x}')\frac{1}{\sin^2\theta}\pi_\phi
	\\
	&=&\frac{\pi_\phi}{\sin^2\theta}\,,
	\\
	\dot\pi_\phi&=&\left\{\pi_\phi,H\right\}=\int d^2x' \left\{ \pi_\phi(\mathbf{x}),\mathcal{H}\right\}
	\\
	&=&\int d^2x' \left\{ \pi_\phi(\mathbf{x}),\frac{1}{2}\sin^2\theta \bm\nabla \phi \cdot \bm \nabla \phi\right\}
	=\int d^2x' \sin^2\theta \sum_i \left\{ \pi_\phi(\mathbf{x}), \partial_i \phi\right\} \partial_i \phi
	\\
	&=&\int d^2x' \sin^2\theta \sum_i (-1)\partial_i \delta(\mathbf{x}-\mathbf{x}') \partial_i \phi
	=\int d^2x'  \sum_i \delta(\mathbf{x}-\mathbf{x}') \partial_i \left(\sin^2\theta\partial_i  \phi\right) 
	\\
	&=&\int d^2x' \delta(\mathbf{x}-\mathbf{x}')\bm\nabla\cdot\left(\sin^2\theta \bm\nabla \phi\right)
	\\
	&=&\bm\nabla\cdot\left(\sin^2\theta \bm\nabla \phi\right)\,. \label{eqpiphi}
\end{eqnarray}
Since $\pi_\phi=\sin^2\theta \dot \phi$, Eq.~\eqref{eqpiphi} is identical to Eq.~\eqref{spincurrent}.

\section{Second variation of the Lagrangian}\label{applag}
The N\'eel vector is an unit vector on the sphere $S^2$.
Thus the N\'eel vector can be written as a rotation of a constant unit vector
\begin{equation}
	\mathbf{n}(\mathbf{x},t)=\mathcal{R}(\mathbf{x},t)\mathbf{e}\,.
\end{equation}
where the rotation matrix is defined by
\begin{equation}
	\mathcal{R}=e^{\phi(\mathbf{x},t) L_z} e^{\theta(\mathbf{x},t) L_y}\,,
\end{equation}
and the generators are
\begin{equation}\label{rotge}
	L_x=
	\begin{pmatrix}
		0 & 0 & 0
		\\
		0 & 0 & -1
		\\
		0 & 1 & 0
	\end{pmatrix}\,,\quad
	L_y=
	\begin{pmatrix}
		0 & 0 & 1
		\\
		0 & 0 & 0
		\\
		-1 & 0 & 0
	\end{pmatrix}\,,\quad
	L_z=
	\begin{pmatrix}
		0 & -1 & 0
		\\
		1 & 0 & 0
		\\
		0 & 0 & 0
	\end{pmatrix}\,.
\end{equation}
With this representation, derivatives of the N\'eel vector is written as
\begin{eqnarray}
	\partial_\mu \mathbf{n} &=& \left[ \left(\partial_\mu \mathcal{R}\right) + \mathcal{R} \partial_\mu\right] \mathbf{e}
	\\
	&=&\mathcal{R} \left[ \mathcal{R}^{-1}\left(\partial_\mu \mathcal{R}\right)+\partial_\mu\right]\mathbf{e}
	\\
	&\equiv&\mathcal{R}\left[\mathcal{A}_\mu+\partial_\mu\right]\mathbf{e}\,. \label{cov}
\end{eqnarray}
The Lagrangian density of the Heisenberg antiferromagnet is given by
\begin{equation}
	\mathcal{L}=\frac{1}{2}\partial_\mu \mathbf{n} \cdot \partial^\mu\mathbf{n}\,, \label{lag}
\end{equation}
and this is called the O(3) sigma model. Plugging the Eq.~\eqref{cov} into the Lagrangian~\eqref{lag}, we can obtain the gauged sigma model
\begin{eqnarray}
	\mathcal{L}&=&\frac{1}{2}\partial_\mu \mathbf{n}\cdot \partial^\mu \mathbf{n}
	\\
	&=&\frac{1}{2}\mathcal{R} \left[ \partial_\mu + \mathcal{A}_\mu \right]\mathbf{e}\cdot \mathcal{R}\left[ \partial^\mu + \mathcal{A}^\mu \right]\mathbf{e}
	\\
	&=& \frac{1}{2}\left[ \partial_\mu + \mathcal{A}_\mu \right]\mathbf{e}\cdot \left[ \partial^\mu + \mathcal{A}^\mu \right]\mathbf{e}
	\\
	&=&\frac{1}{2}\mathcal{D}_\mu \mathbf{e}\cdot \mathcal{D}^\mu\mathbf{e}\,,
\end{eqnarray}
with the covariant derivate
\begin{equation}
	\mathcal{D}_\mu \equiv \partial_\mu + \mathcal{A}_\mu\,,
\end{equation}
and the SO(3) gauge field $\mathcal{A}_\mu\in\frak{so}(3)$, where $\frak{so}(3)$ is the Lie-algebra of the Lie-group SO(3). Since the rotation generators~\eqref{rotge} form a basis of $\frak{so}(3)$, we can write the covariant derivative as
\begin{eqnarray}
	\mathcal{D}_\mu &=& \partial_\mu + \mathcal{A}_\mu = \partial_\mu + A_\mu^iL_i\,, 
	\\
	\left[\mathcal{D}_\mu\mathbf{e}\right]^j&=& \partial_\mu \left[\mathbf{e}\right]^j + A_\mu^i \left[L_i\right]\indices{^j_k}\left[\mathbf{e}\right]^k=\partial_\mu \left[\mathbf{e}\right]^j + A_\mu^i \epsilon\indices{_i^j_k}\left[\mathbf{e}\right]^k=\left[\partial_\mu\mathbf{e}-\mathbf{A}_\mu\times\mathbf{e}\right]^j\,,
\end{eqnarray}
where $A_\mu^i$ is the $i$-th component of $\mathbf{A}_\mu$ which is the dual vector of $\mathcal{A}_\mu$ and $\epsilon$ is the Levi-Civita tensor. In the Ref.~\cite{Hill2021}, the author uses $\mathbf{A}_\mu$ as a SO(3) gauge field. The gauge field $\mathcal{A}_\mu$ and its dual $\mathbf{A}_\mu$ have same information and linked by the relation
\begin{equation}
	\left[\mathcal{A}_\mu\right]\indices{^i_j}=\epsilon\indices{^i_j_k}A_\mu^k\,.
\end{equation}

Effects of magnons are captured by the expansion of $\mathbf{e}\approx \hat{z}+\delta \mathbf{e}$, where $\delta \mathbf{e} = \delta e_1 \hat{x} + \delta e_2 \hat{y},\,\lvert\delta\mathbf{e}\rvert\ll1$. We will obtain the Lagrangian of a magnon via the second order variation with respect to $\delta \mathbf{e}$.
The expansion of the Lagrangian is written as
\begin{eqnarray}
	2\mathcal{L}&=&\mathcal{D}_\mu \mathbf{e} \cdot \mathcal{D}^\mu \mathbf{e}
	\\
	&=&\mathcal{A}_\mu \mathbf{e} \cdot \mathcal{A}^\mu \mathbf{e} 
	+
	\partial_\mu \mathbf{e} \cdot \partial^\mu \mathbf{e}
	+
	2\mathcal{A}_\mu \mathbf{e}\cdot \partial^\mu\mathbf{e}
	\\
	&=&
	\mathcal{A}_\mu \hat{z}\cdot \mathcal{A}^\mu \hat{z} 
	+
	\mathcal{A}_\mu \delta\mathbf{e} \cdot \mathcal{A}^\mu \delta\mathbf{e} 
	+
	2\mathcal{A}_\mu \hat{z} \cdot \mathcal{A}^\mu \delta\mathbf{e} 
	+\partial_\mu\delta \mathbf{e} \cdot \partial^\mu \delta\mathbf{e}
	+
	2\mathcal{A}_\mu \delta\mathbf{e}\cdot \partial^\mu\delta\mathbf{e}
	+
	2\mathcal{A}_\mu \hat{z}\cdot \partial^\mu\delta\mathbf{e}\,.
\end{eqnarray}
The zeroth order term is the Lagrangian without magnon and the first order terms vanish due to the stationary action principle. Now, the Lagrangian of magnon(spin wave) is obtained
\begin{eqnarray}
	2\mathcal{L}_\text{SW}&=&\partial_\mu\delta \mathbf{e} \cdot \partial^\mu \delta\mathbf{e}
	+
	\mathcal{A}_\mu \delta\mathbf{e} \cdot \mathcal{A}^\mu \delta\mathbf{e} 
	+
	2\mathcal{A}_\mu \delta\mathbf{e}\cdot \partial^\mu\delta\mathbf{e}
	\\
	&=&
	\partial_\mu\delta \mathbf{e} \cdot \partial^\mu \delta\mathbf{e}
	+
	\left[
	\left[\mathcal{A}_\mu\right]\indices{^1_2}\delta e_2 \hat{x} 
	+\left[\mathcal{A}_\mu\right]\indices{^2_1}\delta e_1 \hat{y} 
	+\left(\left[\mathcal{A}_\mu\right]\indices{^3_1}\delta e_1 
	+\left[\mathcal{A}_\mu\right]\indices{^3_2}\delta e_2
	\right)\hat{z}
	\right]
	\cdot \mathcal{A}^\mu \delta\mathbf{e} 
	+
	2\mathcal{A}_\mu \delta\mathbf{e}\cdot \partial^\mu\delta\mathbf{e}
	\\
	&=&\partial_\mu\delta \mathbf{e} \cdot \partial^\mu \delta\mathbf{e}
	+
	\left[\mathcal{A}_\mu\right]\indices{^1_2}\left[\mathcal{A}^\mu\right]\indices{^1_2}\delta e_2\delta e_2
	+
	\left[\mathcal{A}_\mu\right]\indices{^1_2}\left[\mathcal{A}^\mu\right]\indices{^1_2}\delta e_1\delta e_1
	\\
	&&+
	\left(\left[\mathcal{A}_\mu\right]\indices{^3_1}\delta e_1 
	+\left[\mathcal{A}_\mu\right]\indices{^3_2}\delta e_2
	\right)\left(\left[\mathcal{A}^\mu\right]\indices{^3_1}\delta e_1 
	+\left[\mathcal{A}^\mu\right]\indices{^3_2}\delta e_2
	\right)
	\\
	&&+2\left[\mathcal{A}_\mu\right]\indices{^1_2}\delta e_2 \partial^\mu \delta e_1 
	-2\left[\mathcal{A}_\mu\right]\indices{^1_2}\delta e_1 \partial^\mu \delta e_2 
	\\
	&=&
	\partial_\mu\delta \mathbf{e} \cdot \partial^\mu \delta\mathbf{e}
	+
	a_\mu a^\mu \left(\delta e_1\delta e_1+\delta e_2\delta e_2\right)
	+
	2a^\mu\left(\delta e_2\partial_\mu \delta e_1 - \delta e_1 \partial_\mu \delta e_2\right)
	\\
	&&+\left(\left[\mathcal{A}_\mu\right]\indices{^3_1}\delta e_1 
	+\left[\mathcal{A}_\mu\right]\indices{^3_2}\delta e_2
	\right)\left(\left[\mathcal{A}^\mu\right]\indices{^3_1}\delta e_1 
	+\left[\mathcal{A}^\mu\right]\indices{^3_2}\delta e_2
	\right)\,.
\end{eqnarray}
Here, we define the texture induced gauge field by the longitudinal component of $\mathbf{A}_\mu$
\begin{equation}
	a_\mu \equiv \left[\mathcal{A}_\mu\right]\indices{^1_2}=A_\mu^3\,,
\end{equation}
which is dominant for the dynamics of high-energy magnons.
Let's introduce a magnon wavefunction $\Psi_q\equiv \delta e_1 -q i \delta e_2 $, then the Lagrangian is written as
\begin{eqnarray}
	2\mathcal{L}_\text{SW}
	&=&
	\partial_\mu \Psi_q^* \partial^\mu \Psi_q+a_\mu a^\mu \Psi_q^* \Psi_q-iq a^\mu\left(\Psi_q^* \partial_\mu \Psi_q -\Psi_q \partial_\mu \Psi_q^*\right)
	\\
	&&+\left(\left[\mathcal{A}_\mu\right]\indices{^3_1}\delta e_1 
	+\left[\mathcal{A}_\mu\right]\indices{^3_2}\delta e_2
	\right)\left(\left[\mathcal{A}^\mu\right]\indices{^3_1}\delta e_1 
	+\left[\mathcal{A}^\mu\right]\indices{^3_2}\delta e_2
	\right)\,,
	\\
	&=&\left(D_\mu \Psi_q\right)^*\left(D^\mu\Psi_q\right)+\left(\left[\mathcal{A}_\mu\right]\indices{^3_1}\delta e_1 
	+\left[\mathcal{A}_\mu\right]\indices{^3_2}\delta e_2
	\right)\left(\left[\mathcal{A}^\mu\right]\indices{^3_1}\delta e_1 
	+\left[\mathcal{A}^\mu\right]\indices{^3_2}\delta e_2
	\right)\,, \label{coeq}
\end{eqnarray}
where the covariant derivative is defined as
\begin{equation}
	D_\mu \equiv \partial_\mu +i q a_\mu\,.
\end{equation}
See the expression~\eqref{coeq}, the covariant derivative term induces the $k_\mu$ term while the last term does not. Here, we consider the high-energy magnon limit which means that the background texture varies slowly and this is equivalent to the adiabatic approximation. In the high-energy-magnon limit, the derivative term in the Lagrangian is dominant thereby we can neglect the last term. Hence the Lagrangian of magnon is given by
\begin{equation}
	\mathcal{L}_\text{SW}=\frac{1}{2}\left(D_\mu \Psi_q\right)^*\left(D^\mu\Psi_q\right)\,.
\end{equation}
If we add the anisotropy term to the Hamiltonian, then the length scale of field configurations is given by
\begin{equation}
	\lambda = \sqrt{\frac{A}{K}}\,,
\end{equation}
where $A$ and $K$ are exchange and anisotropy constants. For the adiabatic approximation, $A\gg K$ and this is called the exchange approximation. Thus, in this limit, we discard the anisotropy term in the Lagrangian, even if the system has an anisotropy.

Let $\{ \bm\varepsilon_i\}$ be a local frame co-rotating with the spin texture and $\{ \mathbf{e}_i\}$ be a fixed global frame. The twist of frame is given by 
\begin{eqnarray}
	\bm\varepsilon_i \cdot \partial_\mu\bm\varepsilon_j&=&\mathcal{R} \mathbf{e}_i \cdot \partial_\mu \mathcal{R} \mathbf{e}_j
	\\
	&=&\mathbf{e}_i \cdot \mathcal{R}^{-1} \partial_\mu \mathcal{R} \mathbf{e}_j
	\\
	&=&\mathbf{e}_i \cdot \mathcal{A}_\mu \mathbf{e}_j
	\\
	&=&\left[\mathcal{A}_\mu \right]_{ij}\,.
\end{eqnarray}
The local rotation of a basis $\{\varepsilon_1, \varepsilon_2\}$ endows a local phase rotation of the complex scalar $\Psi_q$. Thus, naturally, $\left[\mathcal{A}_\mu\right]_{12}=A_\mu^3$ becomes U(1) gauge field for the complex scalar $\Psi_q$ and generates the emergent electromagnetism.

\end{widetext}

\bibliography{/Users/seungho/Documents/zot_library.bib}

\begin{thebibliography}{90}%
\makeatletter
\providecommand \@ifxundefined [1]{%
 \@ifx{#1\undefined}
}%
\providecommand \@ifnum [1]{%
 \ifnum #1\expandafter \@firstoftwo
 \else \expandafter \@secondoftwo
 \fi
}%
\providecommand \@ifx [1]{%
 \ifx #1\expandafter \@firstoftwo
 \else \expandafter \@secondoftwo
 \fi
}%
\providecommand \natexlab [1]{#1}%
\providecommand \enquote  [1]{``#1''}%
\providecommand \bibnamefont  [1]{#1}%
\providecommand \bibfnamefont [1]{#1}%
\providecommand \citenamefont [1]{#1}%
\providecommand \href@noop [0]{\@secondoftwo}%
\providecommand \href [0]{\begingroup \@sanitize@url \@href}%
\providecommand \@href[1]{\@@startlink{#1}\@@href}%
\providecommand \@@href[1]{\endgroup#1\@@endlink}%
\providecommand \@sanitize@url [0]{\catcode `\\12\catcode `\$12\catcode
  `\&12\catcode `\#12\catcode `\^12\catcode `\_12\catcode `\%12\relax}%
\providecommand \@@startlink[1]{}%
\providecommand \@@endlink[0]{}%
\providecommand \url  [0]{\begingroup\@sanitize@url \@url }%
\providecommand \@url [1]{\endgroup\@href {#1}{\urlprefix }}%
\providecommand \urlprefix  [0]{URL }%
\providecommand \Eprint [0]{\href }%
\providecommand \doibase [0]{https://doi.org/}%
\providecommand \selectlanguage [0]{\@gobble}%
\providecommand \bibinfo  [0]{\@secondoftwo}%
\providecommand \bibfield  [0]{\@secondoftwo}%
\providecommand \translation [1]{[#1]}%
\providecommand \BibitemOpen [0]{}%
\providecommand \bibitemStop [0]{}%
\providecommand \bibitemNoStop [0]{.\EOS\space}%
\providecommand \EOS [0]{\spacefactor3000\relax}%
\providecommand \BibitemShut  [1]{\csname bibitem#1\endcsname}%
\let\auto@bib@innerbib\@empty
%</preamble>
\bibitem [{\citenamefont {Baltz}\ \emph {et~al.}(2018)\citenamefont {Baltz},
  \citenamefont {Manchon}, \citenamefont {Tsoi}, \citenamefont {Moriyama},
  \citenamefont {Ono},\ and\ \citenamefont {Tserkovnyak}}]{Baltz2018}%
  \BibitemOpen
  \bibfield  {author} {\bibinfo {author} {\bibfnamefont {V.}~\bibnamefont
  {Baltz}}, \bibinfo {author} {\bibfnamefont {A.}~\bibnamefont {Manchon}},
  \bibinfo {author} {\bibfnamefont {M.}~\bibnamefont {Tsoi}}, \bibinfo {author}
  {\bibfnamefont {T.}~\bibnamefont {Moriyama}}, \bibinfo {author}
  {\bibfnamefont {T.}~\bibnamefont {Ono}},\ and\ \bibinfo {author}
  {\bibfnamefont {Y.}~\bibnamefont {Tserkovnyak}},\ }\bibfield  {title}
  {\bibinfo {title} {Antiferromagnetic spintronics},\ }\href
  {https://doi.org/10.1103/RevModPhys.90.015005} {\bibfield  {journal}
  {\bibinfo  {journal} {Rev. Mod. Phys.}\ }\textbf {\bibinfo {volume} {90}},\
  \bibinfo {pages} {015005} (\bibinfo {year} {2018})}\BibitemShut {NoStop}%
\bibitem [{\citenamefont {Jungwirth}\ \emph {et~al.}(2016)\citenamefont
  {Jungwirth}, \citenamefont {Marti}, \citenamefont {Wadley},\ and\
  \citenamefont {Wunderlich}}]{Jungwirth2016}%
  \BibitemOpen
  \bibfield  {author} {\bibinfo {author} {\bibfnamefont {T.}~\bibnamefont
  {Jungwirth}}, \bibinfo {author} {\bibfnamefont {X.}~\bibnamefont {Marti}},
  \bibinfo {author} {\bibfnamefont {P.}~\bibnamefont {Wadley}},\ and\ \bibinfo
  {author} {\bibfnamefont {J.}~\bibnamefont {Wunderlich}},\ }\bibfield  {title}
  {\bibinfo {title} {Antiferromagnetic spintronics},\ }\href
  {https://doi.org/10.1038/nnano.2016.18} {\bibfield  {journal} {\bibinfo
  {journal} {Nat. Nanotechnol.}\ }\textbf {\bibinfo {volume} {11}},\ \bibinfo
  {pages} {231} (\bibinfo {year} {2016})}\BibitemShut {NoStop}%
\bibitem [{\citenamefont {Cheng}\ \emph {et~al.}(2015)\citenamefont {Cheng},
  \citenamefont {Daniels}, \citenamefont {Zhu},\ and\ \citenamefont
  {Xiao}}]{Cheng2015}%
  \BibitemOpen
  \bibfield  {author} {\bibinfo {author} {\bibfnamefont {R.}~\bibnamefont
  {Cheng}}, \bibinfo {author} {\bibfnamefont {M.~W.}\ \bibnamefont {Daniels}},
  \bibinfo {author} {\bibfnamefont {J.-G.}\ \bibnamefont {Zhu}},\ and\ \bibinfo
  {author} {\bibfnamefont {D.}~\bibnamefont {Xiao}},\ }\bibfield  {title}
  {\bibinfo {title} {Ultrafast switching of antiferromagnets via spin-transfer
  torque},\ }\href {https://doi.org/10.1103/PhysRevB.91.064423} {\bibfield
  {journal} {\bibinfo  {journal} {Phys. Rev. B}\ }\textbf {\bibinfo {volume}
  {91}},\ \bibinfo {pages} {064423} (\bibinfo {year} {2015})}\BibitemShut
  {NoStop}%
\bibitem [{\citenamefont {Khoshlahni}\ \emph {et~al.}(2019)\citenamefont
  {Khoshlahni}, \citenamefont {Qaiumzadeh}, \citenamefont {Bergman},\ and\
  \citenamefont {Brataas}}]{Khoshlahni2019}%
  \BibitemOpen
  \bibfield  {author} {\bibinfo {author} {\bibfnamefont {R.}~\bibnamefont
  {Khoshlahni}}, \bibinfo {author} {\bibfnamefont {A.}~\bibnamefont
  {Qaiumzadeh}}, \bibinfo {author} {\bibfnamefont {A.}~\bibnamefont
  {Bergman}},\ and\ \bibinfo {author} {\bibfnamefont {A.}~\bibnamefont
  {Brataas}},\ }\bibfield  {title} {\bibinfo {title} {Ultrafast generation and
  dynamics of isolated skyrmions in antiferromagnetic insulators},\ }\href
  {https://doi.org/10.1103/PhysRevB.99.054423} {\bibfield  {journal} {\bibinfo
  {journal} {Phys. Rev. B}\ }\textbf {\bibinfo {volume} {99}},\ \bibinfo
  {pages} {054423} (\bibinfo {year} {2019})}\BibitemShut {NoStop}%
\bibitem [{\citenamefont {Gongyo}\ \emph {et~al.}(2016)\citenamefont {Gongyo},
  \citenamefont {Kikuchi}, \citenamefont {Hyodo},\ and\ \citenamefont
  {Kunihiro}}]{Gongyo2016}%
  \BibitemOpen
  \bibfield  {author} {\bibinfo {author} {\bibfnamefont {S.}~\bibnamefont
  {Gongyo}}, \bibinfo {author} {\bibfnamefont {Y.}~\bibnamefont {Kikuchi}},
  \bibinfo {author} {\bibfnamefont {T.}~\bibnamefont {Hyodo}},\ and\ \bibinfo
  {author} {\bibfnamefont {T.}~\bibnamefont {Kunihiro}},\ }\bibfield  {title}
  {\bibinfo {title} {Effective field theory and the scattering process for
  magnons in ferromagnets, antiferromagnets, and ferrimagnets},\ }\href
  {https://doi.org/10.1093/ptep/ptw095} {\bibfield  {journal} {\bibinfo
  {journal} {Prog. Theor. Exp. Phys.}\ }\textbf {\bibinfo {volume} {2016}},\
  \bibinfo {pages} {083B01} (\bibinfo {year} {2016})}\BibitemShut {NoStop}%
\bibitem [{\citenamefont {Rezende}\ \emph {et~al.}(2019)\citenamefont
  {Rezende}, \citenamefont {Azevedo},\ and\ \citenamefont
  {{Rodr{\'i}guez-Su{\'a}rez}}}]{Rezende2019}%
  \BibitemOpen
  \bibfield  {author} {\bibinfo {author} {\bibfnamefont {S.~M.}\ \bibnamefont
  {Rezende}}, \bibinfo {author} {\bibfnamefont {A.}~\bibnamefont {Azevedo}},\
  and\ \bibinfo {author} {\bibfnamefont {R.~L.}\ \bibnamefont
  {{Rodr{\'i}guez-Su{\'a}rez}}},\ }\bibfield  {title} {\bibinfo {title}
  {Introduction to antiferromagnetic magnons},\ }\href
  {https://doi.org/10.1063/1.5109132} {\bibfield  {journal} {\bibinfo
  {journal} {J. Appl. Phys.}\ }\textbf {\bibinfo {volume} {126}},\ \bibinfo
  {pages} {151101} (\bibinfo {year} {2019})}\BibitemShut {NoStop}%
\bibitem [{\citenamefont {Hongo}\ \emph {et~al.}(2021)\citenamefont {Hongo},
  \citenamefont {Fujimori}, \citenamefont {Misumi}, \citenamefont {Nitta},\
  and\ \citenamefont {Sakai}}]{Hongo2021b}%
  \BibitemOpen
  \bibfield  {author} {\bibinfo {author} {\bibfnamefont {M.}~\bibnamefont
  {Hongo}}, \bibinfo {author} {\bibfnamefont {T.}~\bibnamefont {Fujimori}},
  \bibinfo {author} {\bibfnamefont {T.}~\bibnamefont {Misumi}}, \bibinfo
  {author} {\bibfnamefont {M.}~\bibnamefont {Nitta}},\ and\ \bibinfo {author}
  {\bibfnamefont {N.}~\bibnamefont {Sakai}},\ }\bibfield  {title} {\bibinfo
  {title} {Effective field theory of magnons: {{Chiral}} magnets and the
  {{Schwinger}} mechanism},\ }\href
  {https://doi.org/10.1103/PhysRevB.104.134403} {\bibfield  {journal} {\bibinfo
   {journal} {Phys. Rev. B}\ }\textbf {\bibinfo {volume} {104}},\ \bibinfo
  {pages} {134403} (\bibinfo {year} {2021})}\BibitemShut {NoStop}%
\bibitem [{\citenamefont {Wang}\ and\ \citenamefont {Wang}(2021)}]{Wang2021}%
  \BibitemOpen
  \bibfield  {author} {\bibinfo {author} {\bibfnamefont {X.~S.}\ \bibnamefont
  {Wang}}\ and\ \bibinfo {author} {\bibfnamefont {X.~R.}\ \bibnamefont
  {Wang}},\ }\bibfield  {title} {\bibinfo {title} {Topological magnonics},\
  }\href {https://doi.org/10.1063/5.0041781} {\bibfield  {journal} {\bibinfo
  {journal} {J. Appl. Phys.}\ }\textbf {\bibinfo {volume} {129}},\ \bibinfo
  {pages} {151101} (\bibinfo {year} {2021})}\BibitemShut {NoStop}%
\bibitem [{Note1()}]{Note1}%
  \BibitemOpen
  \bibinfo {note} {In our work, we use two terms “spin wave” and
  “magnon” interchangeably with emphasis on wave-like and particle-like
  properties of magnetic excitations, respectively.}\BibitemShut {Stop}%
\bibitem [{\citenamefont {Chumak}\ \emph {et~al.}(2015)\citenamefont {Chumak},
  \citenamefont {Vasyuchka}, \citenamefont {Serga},\ and\ \citenamefont
  {Hillebrands}}]{Chumak2015}%
  \BibitemOpen
  \bibfield  {author} {\bibinfo {author} {\bibfnamefont {A.~V.}\ \bibnamefont
  {Chumak}}, \bibinfo {author} {\bibfnamefont {V.~I.}\ \bibnamefont
  {Vasyuchka}}, \bibinfo {author} {\bibfnamefont {A.~A.}\ \bibnamefont
  {Serga}},\ and\ \bibinfo {author} {\bibfnamefont {B.}~\bibnamefont
  {Hillebrands}},\ }\bibfield  {title} {\bibinfo {title} {Magnon spintronics},\
  }\href {https://doi.org/10.1038/nphys3347} {\bibfield  {journal} {\bibinfo
  {journal} {Nat. Phys.}\ }\textbf {\bibinfo {volume} {11}},\ \bibinfo {pages}
  {453} (\bibinfo {year} {2015})}\BibitemShut {NoStop}%
\bibitem [{\citenamefont {Barman}\ \emph {et~al.}(2021)\citenamefont {Barman},
  \citenamefont {Gubbiotti}, \citenamefont {Ladak}, \citenamefont {Adeyeye},
  \citenamefont {Krawczyk}, \citenamefont {Gr{\"a}fe}, \citenamefont
  {Adelmann}, \citenamefont {Cotofana}, \citenamefont {Naeemi}, \citenamefont
  {Vasyuchka}, \citenamefont {Hillebrands}, \citenamefont {Nikitov},
  \citenamefont {Yu}, \citenamefont {Grundler}, \citenamefont {Sadovnikov},
  \citenamefont {Grachev}, \citenamefont {Sheshukova}, \citenamefont
  {Duquesne}, \citenamefont {Marangolo}, \citenamefont {Csaba}, \citenamefont
  {Porod}, \citenamefont {Demidov}, \citenamefont {Urazhdin}, \citenamefont
  {Demokritov}, \citenamefont {Albisetti}, \citenamefont {Petti}, \citenamefont
  {Bertacco}, \citenamefont {Schultheiss}, \citenamefont {Kruglyak},
  \citenamefont {Poimanov}, \citenamefont {Sahoo}, \citenamefont {Sinha},
  \citenamefont {Yang}, \citenamefont {M{\"u}nzenberg}, \citenamefont
  {Moriyama}, \citenamefont {Mizukami}, \citenamefont {Landeros}, \citenamefont
  {Gallardo}, \citenamefont {Carlotti}, \citenamefont {Kim}, \citenamefont
  {Stamps}, \citenamefont {Camley}, \citenamefont {Rana}, \citenamefont
  {Otani}, \citenamefont {Yu}, \citenamefont {Yu}, \citenamefont {Bauer},
  \citenamefont {Back}, \citenamefont {Uhrig}, \citenamefont {Dobrovolskiy},
  \citenamefont {Budinska}, \citenamefont {Qin}, \citenamefont {{van Dijken}},
  \citenamefont {Chumak}, \citenamefont {Khitun}, \citenamefont {Nikonov},
  \citenamefont {Young}, \citenamefont {Zingsem},\ and\ \citenamefont
  {Winklhofer}}]{Barman2021}%
  \BibitemOpen
  \bibfield  {author} {\bibinfo {author} {\bibfnamefont {A.}~\bibnamefont
  {Barman}}, \bibinfo {author} {\bibfnamefont {G.}~\bibnamefont {Gubbiotti}},
  \bibinfo {author} {\bibfnamefont {S.}~\bibnamefont {Ladak}}, \bibinfo
  {author} {\bibfnamefont {A.~O.}\ \bibnamefont {Adeyeye}}, \bibinfo {author}
  {\bibfnamefont {M.}~\bibnamefont {Krawczyk}}, \bibinfo {author}
  {\bibfnamefont {J.}~\bibnamefont {Gr{\"a}fe}}, \bibinfo {author}
  {\bibfnamefont {C.}~\bibnamefont {Adelmann}}, \bibinfo {author}
  {\bibfnamefont {S.}~\bibnamefont {Cotofana}}, \bibinfo {author}
  {\bibfnamefont {A.}~\bibnamefont {Naeemi}}, \bibinfo {author} {\bibfnamefont
  {V.~I.}\ \bibnamefont {Vasyuchka}}, \bibinfo {author} {\bibfnamefont
  {B.}~\bibnamefont {Hillebrands}}, \bibinfo {author} {\bibfnamefont {S.~A.}\
  \bibnamefont {Nikitov}}, \bibinfo {author} {\bibfnamefont {H.}~\bibnamefont
  {Yu}}, \bibinfo {author} {\bibfnamefont {D.}~\bibnamefont {Grundler}},
  \bibinfo {author} {\bibfnamefont {A.~V.}\ \bibnamefont {Sadovnikov}},
  \bibinfo {author} {\bibfnamefont {A.~A.}\ \bibnamefont {Grachev}}, \bibinfo
  {author} {\bibfnamefont {S.~E.}\ \bibnamefont {Sheshukova}}, \bibinfo
  {author} {\bibfnamefont {J.-Y.}\ \bibnamefont {Duquesne}}, \bibinfo {author}
  {\bibfnamefont {M.}~\bibnamefont {Marangolo}}, \bibinfo {author}
  {\bibfnamefont {G.}~\bibnamefont {Csaba}}, \bibinfo {author} {\bibfnamefont
  {W.}~\bibnamefont {Porod}}, \bibinfo {author} {\bibfnamefont {V.~E.}\
  \bibnamefont {Demidov}}, \bibinfo {author} {\bibfnamefont {S.}~\bibnamefont
  {Urazhdin}}, \bibinfo {author} {\bibfnamefont {S.~O.}\ \bibnamefont
  {Demokritov}}, \bibinfo {author} {\bibfnamefont {E.}~\bibnamefont
  {Albisetti}}, \bibinfo {author} {\bibfnamefont {D.}~\bibnamefont {Petti}},
  \bibinfo {author} {\bibfnamefont {R.}~\bibnamefont {Bertacco}}, \bibinfo
  {author} {\bibfnamefont {H.}~\bibnamefont {Schultheiss}}, \bibinfo {author}
  {\bibfnamefont {V.~V.}\ \bibnamefont {Kruglyak}}, \bibinfo {author}
  {\bibfnamefont {V.~D.}\ \bibnamefont {Poimanov}}, \bibinfo {author}
  {\bibfnamefont {S.}~\bibnamefont {Sahoo}}, \bibinfo {author} {\bibfnamefont
  {J.}~\bibnamefont {Sinha}}, \bibinfo {author} {\bibfnamefont
  {H.}~\bibnamefont {Yang}}, \bibinfo {author} {\bibfnamefont {M.}~\bibnamefont
  {M{\"u}nzenberg}}, \bibinfo {author} {\bibfnamefont {T.}~\bibnamefont
  {Moriyama}}, \bibinfo {author} {\bibfnamefont {S.}~\bibnamefont {Mizukami}},
  \bibinfo {author} {\bibfnamefont {P.}~\bibnamefont {Landeros}}, \bibinfo
  {author} {\bibfnamefont {R.~A.}\ \bibnamefont {Gallardo}}, \bibinfo {author}
  {\bibfnamefont {G.}~\bibnamefont {Carlotti}}, \bibinfo {author}
  {\bibfnamefont {J.-V.}\ \bibnamefont {Kim}}, \bibinfo {author} {\bibfnamefont
  {R.~L.}\ \bibnamefont {Stamps}}, \bibinfo {author} {\bibfnamefont {R.~E.}\
  \bibnamefont {Camley}}, \bibinfo {author} {\bibfnamefont {B.}~\bibnamefont
  {Rana}}, \bibinfo {author} {\bibfnamefont {Y.}~\bibnamefont {Otani}},
  \bibinfo {author} {\bibfnamefont {W.}~\bibnamefont {Yu}}, \bibinfo {author}
  {\bibfnamefont {T.}~\bibnamefont {Yu}}, \bibinfo {author} {\bibfnamefont
  {G.~E.~W.}\ \bibnamefont {Bauer}}, \bibinfo {author} {\bibfnamefont
  {C.}~\bibnamefont {Back}}, \bibinfo {author} {\bibfnamefont {G.~S.}\
  \bibnamefont {Uhrig}}, \bibinfo {author} {\bibfnamefont {O.~V.}\ \bibnamefont
  {Dobrovolskiy}}, \bibinfo {author} {\bibfnamefont {B.}~\bibnamefont
  {Budinska}}, \bibinfo {author} {\bibfnamefont {H.}~\bibnamefont {Qin}},
  \bibinfo {author} {\bibfnamefont {S.}~\bibnamefont {{van Dijken}}}, \bibinfo
  {author} {\bibfnamefont {A.~V.}\ \bibnamefont {Chumak}}, \bibinfo {author}
  {\bibfnamefont {A.}~\bibnamefont {Khitun}}, \bibinfo {author} {\bibfnamefont
  {D.~E.}\ \bibnamefont {Nikonov}}, \bibinfo {author} {\bibfnamefont {I.~A.}\
  \bibnamefont {Young}}, \bibinfo {author} {\bibfnamefont {B.~W.}\ \bibnamefont
  {Zingsem}},\ and\ \bibinfo {author} {\bibfnamefont {M.}~\bibnamefont
  {Winklhofer}},\ }\bibfield  {title} {\bibinfo {title} {The 2021 {{Magnonics
  Roadmap}}},\ }\href {https://doi.org/10.1088/1361-648X/abec1a} {\bibfield
  {journal} {\bibinfo  {journal} {J. Phys. Condens. Matter}\ }\textbf {\bibinfo
  {volume} {33}},\ \bibinfo {pages} {413001} (\bibinfo {year}
  {2021})}\BibitemShut {NoStop}%
\bibitem [{\citenamefont {Bauer}\ \emph {et~al.}(2012)\citenamefont {Bauer},
  \citenamefont {Saitoh},\ and\ \citenamefont {{van Wees}}}]{Bauer2012a}%
  \BibitemOpen
  \bibfield  {author} {\bibinfo {author} {\bibfnamefont {G.~E.~W.}\
  \bibnamefont {Bauer}}, \bibinfo {author} {\bibfnamefont {E.}~\bibnamefont
  {Saitoh}},\ and\ \bibinfo {author} {\bibfnamefont {B.~J.}\ \bibnamefont {{van
  Wees}}},\ }\bibfield  {title} {\bibinfo {title} {Spin caloritronics},\ }\href
  {https://doi.org/10.1038/nmat3301} {\bibfield  {journal} {\bibinfo  {journal}
  {Nat. Mater.}\ }\textbf {\bibinfo {volume} {11}},\ \bibinfo {pages} {391}
  (\bibinfo {year} {2012})}\BibitemShut {NoStop}%
\bibitem [{\citenamefont {Manton}\ and\ \citenamefont
  {Sutcliffe}(2004)}]{Manton2004a}%
  \BibitemOpen
  \bibfield  {author} {\bibinfo {author} {\bibfnamefont {N.}~\bibnamefont
  {Manton}}\ and\ \bibinfo {author} {\bibfnamefont {P.}~\bibnamefont
  {Sutcliffe}},\ }\href {https://doi.org/10.1017/CBO9780511617034} {\emph
  {\bibinfo {title} {Topological {{Solitons}}}}},\ Cambridge {{Monographs}} on
  {{Mathematical Physics}}\ (\bibinfo  {publisher} {{Cambridge University
  Press}},\ \bibinfo {address} {{Cambridge}},\ \bibinfo {year}
  {2004})\BibitemShut {NoStop}%
\bibitem [{\citenamefont {Vachaspati}(2006)}]{Vachaspati2006}%
  \BibitemOpen
  \bibfield  {author} {\bibinfo {author} {\bibfnamefont {T.}~\bibnamefont
  {Vachaspati}},\ }\href {https://doi.org/10.1017/CBO9780511535192} {\emph
  {\bibinfo {title} {Kinks and {{Domain Walls}}: {{An Introduction}} to
  {{Classical}} and {{Quantum Solitons}}}}}\ (\bibinfo  {publisher} {{Cambridge
  University Press}},\ \bibinfo {address} {{Cambridge}},\ \bibinfo {year}
  {2006})\BibitemShut {NoStop}%
\bibitem [{\citenamefont {Weinberg}(2012)}]{Weinberg2012}%
  \BibitemOpen
  \bibfield  {author} {\bibinfo {author} {\bibfnamefont {E.~J.}\ \bibnamefont
  {Weinberg}},\ }\href {https://doi.org/10.1017/CBO9781139017787} {\emph
  {\bibinfo {title} {Classical {{Solutions}} in {{Quantum Field Theory}}:
  {{Solitons}} and {{Instantons}} in {{High Energy Physics}}}}},\ Cambridge
  {{Monographs}} on {{Mathematical Physics}}\ (\bibinfo  {publisher}
  {{Cambridge University Press}},\ \bibinfo {address} {{Cambridge}},\ \bibinfo
  {year} {2012})\BibitemShut {NoStop}%
\bibitem [{\citenamefont {Shnir}(2018)}]{Shnir2018}%
  \BibitemOpen
  \bibfield  {author} {\bibinfo {author} {\bibfnamefont {Y.~M.}\ \bibnamefont
  {Shnir}},\ }\href {https://doi.org/10.1017/9781108555623} {\emph {\bibinfo
  {title} {Topological and {{Non-Topological Solitons}} in {{Scalar Field
  Theories}}}}},\ Cambridge {{Monographs}} on {{Mathematical Physics}}\
  (\bibinfo  {publisher} {{Cambridge University Press}},\ \bibinfo {address}
  {{Cambridge}},\ \bibinfo {year} {2018})\BibitemShut {NoStop}%
\bibitem [{\citenamefont {Nitta}(2022)}]{Nitta2022}%
  \BibitemOpen
  \bibfield  {author} {\bibinfo {author} {\bibfnamefont {M.}~\bibnamefont
  {Nitta}},\ }\bibfield  {title} {\bibinfo {title} {Relations among topological
  solitons},\ }\href {https://doi.org/10.1103/PhysRevD.105.105006} {\bibfield
  {journal} {\bibinfo  {journal} {Phys. Rev. D}\ }\textbf {\bibinfo {volume}
  {105}},\ \bibinfo {pages} {105006} (\bibinfo {year} {2022})}\BibitemShut
  {NoStop}%
\bibitem [{\citenamefont {Vento}(2017)}]{Vento2017}%
  \BibitemOpen
  \bibfield  {author} {\bibinfo {author} {\bibfnamefont {V.}~\bibnamefont
  {Vento}},\ }\bibfield  {title} {\bibinfo {title} {Skyrmions at high
  density},\ }\href {https://doi.org/10.1142/S0218301317400298} {\bibfield
  {journal} {\bibinfo  {journal} {Int. J. Mod. Phys. E}\ }\textbf {\bibinfo
  {volume} {26}},\ \bibinfo {pages} {1740029} (\bibinfo {year}
  {2017})}\BibitemShut {NoStop}%
\bibitem [{\citenamefont {Park}\ \emph {et~al.}(2019)\citenamefont {Park},
  \citenamefont {Paeng},\ and\ \citenamefont {Vento}}]{Park2019a}%
  \BibitemOpen
  \bibfield  {author} {\bibinfo {author} {\bibfnamefont {B.-Y.}\ \bibnamefont
  {Park}}, \bibinfo {author} {\bibfnamefont {W.-G.}\ \bibnamefont {Paeng}},\
  and\ \bibinfo {author} {\bibfnamefont {V.}~\bibnamefont {Vento}},\ }\bibfield
   {title} {\bibinfo {title} {The inhomogeneous phase of dense skyrmion
  matter},\ }\href {https://doi.org/10.1016/j.nuclphysa.2019.06.010} {\bibfield
   {journal} {\bibinfo  {journal} {Nucl. Phys. A}\ }\textbf {\bibinfo {volume}
  {989}},\ \bibinfo {pages} {231} (\bibinfo {year} {2019})}\BibitemShut
  {NoStop}%
\bibitem [{\citenamefont {Ackerman}\ \emph {et~al.}(2015)\citenamefont
  {Ackerman}, \citenamefont {{van de Lagemaat}},\ and\ \citenamefont
  {Smalyukh}}]{Ackerman2015}%
  \BibitemOpen
  \bibfield  {author} {\bibinfo {author} {\bibfnamefont {P.~J.}\ \bibnamefont
  {Ackerman}}, \bibinfo {author} {\bibfnamefont {J.}~\bibnamefont {{van de
  Lagemaat}}},\ and\ \bibinfo {author} {\bibfnamefont {I.~I.}\ \bibnamefont
  {Smalyukh}},\ }\bibfield  {title} {\bibinfo {title} {Self-assembly and
  electrostriction of arrays and chains of hopfion particles in chiral liquid
  crystals},\ }\href {https://doi.org/10.1038/ncomms7012} {\bibfield  {journal}
  {\bibinfo  {journal} {Nat Commun}\ }\textbf {\bibinfo {volume} {6}},\
  \bibinfo {pages} {6012} (\bibinfo {year} {2015})}\BibitemShut {NoStop}%
\bibitem [{\citenamefont {Ackerman}\ and\ \citenamefont
  {Smalyukh}(2017)}]{Ackerman2017a}%
  \BibitemOpen
  \bibfield  {author} {\bibinfo {author} {\bibfnamefont {P.~J.}\ \bibnamefont
  {Ackerman}}\ and\ \bibinfo {author} {\bibfnamefont {I.~I.}\ \bibnamefont
  {Smalyukh}},\ }\bibfield  {title} {\bibinfo {title} {Diversity of {{Knot
  Solitons}} in {{Liquid Crystals Manifested}} by {{Linking}} of {{Preimages}}
  in {{Torons}} and {{Hopfions}}},\ }\href
  {https://doi.org/10.1103/PhysRevX.7.011006} {\bibfield  {journal} {\bibinfo
  {journal} {Phys. Rev. X}\ }\textbf {\bibinfo {volume} {7}},\ \bibinfo {pages}
  {011006} (\bibinfo {year} {2017})}\BibitemShut {NoStop}%
\bibitem [{\citenamefont {Afghah}\ and\ \citenamefont
  {Selinger}(2017)}]{Afghah2017a}%
  \BibitemOpen
  \bibfield  {author} {\bibinfo {author} {\bibfnamefont {S.}~\bibnamefont
  {Afghah}}\ and\ \bibinfo {author} {\bibfnamefont {J.~V.}\ \bibnamefont
  {Selinger}},\ }\bibfield  {title} {\bibinfo {title} {Theory of helicoids and
  skyrmions in confined cholesteric liquid crystals},\ }\href
  {https://doi.org/10.1103/PhysRevE.96.012708} {\bibfield  {journal} {\bibinfo
  {journal} {Phys. Rev. E}\ }\textbf {\bibinfo {volume} {96}},\ \bibinfo
  {pages} {012708} (\bibinfo {year} {2017})}\BibitemShut {NoStop}%
\bibitem [{\citenamefont {Long}\ \emph {et~al.}(2021)\citenamefont {Long},
  \citenamefont {Tang}, \citenamefont {Selinger},\ and\ \citenamefont
  {Selinger}}]{Long2021a}%
  \BibitemOpen
  \bibfield  {author} {\bibinfo {author} {\bibfnamefont {C.}~\bibnamefont
  {Long}}, \bibinfo {author} {\bibfnamefont {X.}~\bibnamefont {Tang}}, \bibinfo
  {author} {\bibfnamefont {R.~L.~B.}\ \bibnamefont {Selinger}},\ and\ \bibinfo
  {author} {\bibfnamefont {J.~V.}\ \bibnamefont {Selinger}},\ }\bibfield
  {title} {\bibinfo {title} {Geometry and mechanics of disclination lines in
  {{3D}} nematic liquid crystals},\ }\href {https://doi.org/10.1039/D0SM01899F}
  {\bibfield  {journal} {\bibinfo  {journal} {Soft Matter}\ }\textbf {\bibinfo
  {volume} {17}},\ \bibinfo {pages} {2265} (\bibinfo {year}
  {2021})}\BibitemShut {NoStop}%
\bibitem [{\citenamefont {Long}\ and\ \citenamefont
  {Selinger}(2021)}]{Long2021}%
  \BibitemOpen
  \bibfield  {author} {\bibinfo {author} {\bibfnamefont {C.}~\bibnamefont
  {Long}}\ and\ \bibinfo {author} {\bibfnamefont {J.~V.}\ \bibnamefont
  {Selinger}},\ }\bibfield  {title} {\bibinfo {title} {Coarse-grained theory
  for motion of solitons and skyrmions in liquid crystals},\ }\href
  {https://doi.org/10.1039/D1SM01335A} {\bibfield  {journal} {\bibinfo
  {journal} {Soft Matter}\ }\textbf {\bibinfo {volume} {17}},\ \bibinfo {pages}
  {10437} (\bibinfo {year} {2021})}\BibitemShut {NoStop}%
\bibitem [{\citenamefont {Schimming}\ and\ \citenamefont
  {Vi{\~n}als}(2022)}]{Schimming2022}%
  \BibitemOpen
  \bibfield  {author} {\bibinfo {author} {\bibfnamefont {C.~D.}\ \bibnamefont
  {Schimming}}\ and\ \bibinfo {author} {\bibfnamefont {J.}~\bibnamefont
  {Vi{\~n}als}},\ }\bibfield  {title} {\bibinfo {title} {Singularity
  identification for the characterization of topology, geometry, and motion of
  nematic disclination lines},\ }\href {https://doi.org/10.1039/D1SM01584B}
  {\bibfield  {journal} {\bibinfo  {journal} {Soft Matter}\ }\textbf {\bibinfo
  {volume} {18}},\ \bibinfo {pages} {2234} (\bibinfo {year}
  {2022})}\BibitemShut {NoStop}%
\bibitem [{\citenamefont {Schimming}\ and\ \citenamefont
  {Vi{\~n}als}(2020)}]{Schimming2020a}%
  \BibitemOpen
  \bibfield  {author} {\bibinfo {author} {\bibfnamefont {C.~D.}\ \bibnamefont
  {Schimming}}\ and\ \bibinfo {author} {\bibfnamefont {J.}~\bibnamefont
  {Vi{\~n}als}},\ }\bibfield  {title} {\bibinfo {title} {Anisotropic
  disclination cores in nematic liquid crystals modeled by a self-consistent
  molecular field theory},\ }\href
  {https://doi.org/10.1103/PhysRevE.102.010701} {\bibfield  {journal} {\bibinfo
   {journal} {Phys. Rev. E}\ }\textbf {\bibinfo {volume} {102}},\ \bibinfo
  {pages} {010701} (\bibinfo {year} {2020})}\BibitemShut {NoStop}%
\bibitem [{\citenamefont {Parmee}\ \emph {et~al.}(2022)\citenamefont {Parmee},
  \citenamefont {Dennis},\ and\ \citenamefont {Ruostekoski}}]{Parmee2022}%
  \BibitemOpen
  \bibfield  {author} {\bibinfo {author} {\bibfnamefont {C.~D.}\ \bibnamefont
  {Parmee}}, \bibinfo {author} {\bibfnamefont {M.~R.}\ \bibnamefont {Dennis}},\
  and\ \bibinfo {author} {\bibfnamefont {J.}~\bibnamefont {Ruostekoski}},\
  }\bibfield  {title} {\bibinfo {title} {Optical excitations of {{Skyrmions}},
  knotted solitons, and defects in atoms},\ }\href
  {https://doi.org/10.1038/s42005-022-00829-y} {\bibfield  {journal} {\bibinfo
  {journal} {Commun. Phys.}\ }\textbf {\bibinfo {volume} {5}},\ \bibinfo
  {pages} {1} (\bibinfo {year} {2022})}\BibitemShut {NoStop}%
\bibitem [{\citenamefont {Poy}\ \emph {et~al.}(2022)\citenamefont {Poy},
  \citenamefont {Hess}, \citenamefont {Seracuse}, \citenamefont {Paul},
  \citenamefont {{\v Z}umer},\ and\ \citenamefont {Smalyukh}}]{Poy2022}%
  \BibitemOpen
  \bibfield  {author} {\bibinfo {author} {\bibfnamefont {G.}~\bibnamefont
  {Poy}}, \bibinfo {author} {\bibfnamefont {A.~J.}\ \bibnamefont {Hess}},
  \bibinfo {author} {\bibfnamefont {A.~J.}\ \bibnamefont {Seracuse}}, \bibinfo
  {author} {\bibfnamefont {M.}~\bibnamefont {Paul}}, \bibinfo {author}
  {\bibfnamefont {S.}~\bibnamefont {{\v Z}umer}},\ and\ \bibinfo {author}
  {\bibfnamefont {I.~I.}\ \bibnamefont {Smalyukh}},\ }\bibfield  {title}
  {\bibinfo {title} {Interaction and co-assembly of optical and topological
  solitons},\ }\href {https://doi.org/10.1038/s41566-022-01002-1} {\bibfield
  {journal} {\bibinfo  {journal} {Nat. Photon.}\ }\textbf {\bibinfo {volume}
  {16}},\ \bibinfo {pages} {454} (\bibinfo {year} {2022})}\BibitemShut
  {NoStop}%
\bibitem [{\citenamefont {Lin}\ and\ \citenamefont {Hayami}(2016)}]{Lin2016}%
  \BibitemOpen
  \bibfield  {author} {\bibinfo {author} {\bibfnamefont {S.-Z.}\ \bibnamefont
  {Lin}}\ and\ \bibinfo {author} {\bibfnamefont {S.}~\bibnamefont {Hayami}},\
  }\bibfield  {title} {\bibinfo {title} {Ginzburg-{{Landau}} theory for
  skyrmions in inversion-symmetric magnets with competing interactions},\
  }\href {https://doi.org/10.1103/PhysRevB.93.064430} {\bibfield  {journal}
  {\bibinfo  {journal} {Phys. Rev. B}\ }\textbf {\bibinfo {volume} {93}},\
  \bibinfo {pages} {064430} (\bibinfo {year} {2016})}\BibitemShut {NoStop}%
\bibitem [{\citenamefont {Psaroudaki}\ \emph {et~al.}(2017)\citenamefont
  {Psaroudaki}, \citenamefont {Hoffman}, \citenamefont {Klinovaja},\ and\
  \citenamefont {Loss}}]{Psaroudaki2017}%
  \BibitemOpen
  \bibfield  {author} {\bibinfo {author} {\bibfnamefont {C.}~\bibnamefont
  {Psaroudaki}}, \bibinfo {author} {\bibfnamefont {S.}~\bibnamefont {Hoffman}},
  \bibinfo {author} {\bibfnamefont {J.}~\bibnamefont {Klinovaja}},\ and\
  \bibinfo {author} {\bibfnamefont {D.}~\bibnamefont {Loss}},\ }\bibfield
  {title} {\bibinfo {title} {Quantum {{Dynamics}} of {{Skyrmions}} in {{Chiral
  Magnets}}},\ }\href {https://doi.org/10.1103/PhysRevX.7.041045} {\bibfield
  {journal} {\bibinfo  {journal} {Phys. Rev. X}\ }\textbf {\bibinfo {volume}
  {7}},\ \bibinfo {pages} {041045} (\bibinfo {year} {2017})}\BibitemShut
  {NoStop}%
\bibitem [{\citenamefont {D{\'i}az}\ \emph {et~al.}(2019)\citenamefont
  {D{\'i}az}, \citenamefont {Klinovaja},\ and\ \citenamefont
  {Loss}}]{Diaz2019}%
  \BibitemOpen
  \bibfield  {author} {\bibinfo {author} {\bibfnamefont {S.~A.}\ \bibnamefont
  {D{\'i}az}}, \bibinfo {author} {\bibfnamefont {J.}~\bibnamefont
  {Klinovaja}},\ and\ \bibinfo {author} {\bibfnamefont {D.}~\bibnamefont
  {Loss}},\ }\bibfield  {title} {\bibinfo {title} {Topological {{Magnons}} and
  {{Edge States}} in {{Antiferromagnetic Skyrmion Crystals}}},\ }\href
  {https://doi.org/10.1103/PhysRevLett.122.187203} {\bibfield  {journal}
  {\bibinfo  {journal} {Phys. Rev. Lett.}\ }\textbf {\bibinfo {volume} {122}},\
  \bibinfo {pages} {187203} (\bibinfo {year} {2019})}\BibitemShut {NoStop}%
\bibitem [{\citenamefont {D{\'i}az}\ \emph {et~al.}(2020)\citenamefont
  {D{\'i}az}, \citenamefont {Hirosawa}, \citenamefont {Klinovaja},\ and\
  \citenamefont {Loss}}]{Diaz2020a}%
  \BibitemOpen
  \bibfield  {author} {\bibinfo {author} {\bibfnamefont {S.~A.}\ \bibnamefont
  {D{\'i}az}}, \bibinfo {author} {\bibfnamefont {T.}~\bibnamefont {Hirosawa}},
  \bibinfo {author} {\bibfnamefont {J.}~\bibnamefont {Klinovaja}},\ and\
  \bibinfo {author} {\bibfnamefont {D.}~\bibnamefont {Loss}},\ }\bibfield
  {title} {\bibinfo {title} {Chiral magnonic edge states in ferromagnetic
  skyrmion crystals controlled by magnetic fields},\ }\href
  {https://doi.org/10.1103/PhysRevResearch.2.013231} {\bibfield  {journal}
  {\bibinfo  {journal} {Phys. Rev. Res.}\ }\textbf {\bibinfo {volume} {2}},\
  \bibinfo {pages} {013231} (\bibinfo {year} {2020})}\BibitemShut {NoStop}%
\bibitem [{\citenamefont {Psaroudaki}\ and\ \citenamefont
  {Panagopoulos}(2021)}]{Psaroudaki2021}%
  \BibitemOpen
  \bibfield  {author} {\bibinfo {author} {\bibfnamefont {C.}~\bibnamefont
  {Psaroudaki}}\ and\ \bibinfo {author} {\bibfnamefont {C.}~\bibnamefont
  {Panagopoulos}},\ }\bibfield  {title} {\bibinfo {title} {Skyrmion {{Qubits}}:
  {{A New Class}} of {{Quantum Logic Elements Based}} on {{Nanoscale
  Magnetization}}},\ }\href {https://doi.org/10.1103/PhysRevLett.127.067201}
  {\bibfield  {journal} {\bibinfo  {journal} {Phys. Rev. Lett.}\ }\textbf
  {\bibinfo {volume} {127}},\ \bibinfo {pages} {067201} (\bibinfo {year}
  {2021})}\BibitemShut {NoStop}%
\bibitem [{\citenamefont {Hayami}\ \emph {et~al.}(2021)\citenamefont {Hayami},
  \citenamefont {Okubo},\ and\ \citenamefont {Motome}}]{Hayami2021b}%
  \BibitemOpen
  \bibfield  {author} {\bibinfo {author} {\bibfnamefont {S.}~\bibnamefont
  {Hayami}}, \bibinfo {author} {\bibfnamefont {T.}~\bibnamefont {Okubo}},\ and\
  \bibinfo {author} {\bibfnamefont {Y.}~\bibnamefont {Motome}},\ }\bibfield
  {title} {\bibinfo {title} {Phase shift in skyrmion crystals},\ }\href
  {https://doi.org/10.1038/s41467-021-27083-0} {\bibfield  {journal} {\bibinfo
  {journal} {Nat Commun}\ }\textbf {\bibinfo {volume} {12}},\ \bibinfo {pages}
  {6927} (\bibinfo {year} {2021})}\BibitemShut {NoStop}%
\bibitem [{\citenamefont {Psaroudaki}\ and\ \citenamefont
  {Panagopoulos}(2022)}]{Psaroudaki2022}%
  \BibitemOpen
  \bibfield  {author} {\bibinfo {author} {\bibfnamefont {C.}~\bibnamefont
  {Psaroudaki}}\ and\ \bibinfo {author} {\bibfnamefont {C.}~\bibnamefont
  {Panagopoulos}},\ }\bibfield  {title} {\bibinfo {title} {Skyrmion helicity:
  {{Quantization}} and quantum tunneling effects},\ }\href
  {https://doi.org/10.1103/PhysRevB.106.104422} {\bibfield  {journal} {\bibinfo
   {journal} {Phys. Rev. B}\ }\textbf {\bibinfo {volume} {106}},\ \bibinfo
  {pages} {104422} (\bibinfo {year} {2022})}\BibitemShut {NoStop}%
\bibitem [{\citenamefont {Naya}\ \emph {et~al.}(2022)\citenamefont {Naya},
  \citenamefont {Schubring}, \citenamefont {Shifman},\ and\ \citenamefont
  {Wang}}]{Naya2022}%
  \BibitemOpen
  \bibfield  {author} {\bibinfo {author} {\bibfnamefont {C.}~\bibnamefont
  {Naya}}, \bibinfo {author} {\bibfnamefont {D.}~\bibnamefont {Schubring}},
  \bibinfo {author} {\bibfnamefont {M.}~\bibnamefont {Shifman}},\ and\ \bibinfo
  {author} {\bibfnamefont {Z.}~\bibnamefont {Wang}},\ }\bibfield  {title}
  {\bibinfo {title} {Skyrmions and hopfions in three-dimensional frustrated
  magnets},\ }\href {https://doi.org/10.1103/PhysRevB.106.094434} {\bibfield
  {journal} {\bibinfo  {journal} {Phys. Rev. B}\ }\textbf {\bibinfo {volume}
  {106}},\ \bibinfo {pages} {094434} (\bibinfo {year} {2022})}\BibitemShut
  {NoStop}%
\bibitem [{\citenamefont {Hayami}(2022)}]{Hayami2022e}%
  \BibitemOpen
  \bibfield  {author} {\bibinfo {author} {\bibfnamefont {S.}~\bibnamefont
  {Hayami}},\ }\bibfield  {title} {\bibinfo {title} {Ferroaxial moment induced
  by vortex spin texture},\ }\href
  {https://doi.org/10.1103/PhysRevB.106.144402} {\bibfield  {journal} {\bibinfo
   {journal} {Phys. Rev. B}\ }\textbf {\bibinfo {volume} {106}},\ \bibinfo
  {pages} {144402} (\bibinfo {year} {2022})}\BibitemShut {NoStop}%
\bibitem [{\citenamefont {Back}\ \emph {et~al.}(2020)\citenamefont {Back},
  \citenamefont {Cros}, \citenamefont {Ebert}, \citenamefont
  {{Everschor-Sitte}}, \citenamefont {Fert}, \citenamefont {Garst},
  \citenamefont {Ma}, \citenamefont {Mankovsky}, \citenamefont {Monchesky},
  \citenamefont {Mostovoy}, \citenamefont {Nagaosa}, \citenamefont {Parkin},
  \citenamefont {Pfleiderer}, \citenamefont {Reyren}, \citenamefont {Rosch},
  \citenamefont {Taguchi},\ and\ \citenamefont {Tokura}}]{Back2020}%
  \BibitemOpen
  \bibfield  {author} {\bibinfo {author} {\bibfnamefont {C.}~\bibnamefont
  {Back}}, \bibinfo {author} {\bibfnamefont {V.}~\bibnamefont {Cros}}, \bibinfo
  {author} {\bibfnamefont {H.}~\bibnamefont {Ebert}}, \bibinfo {author}
  {\bibfnamefont {K.}~\bibnamefont {{Everschor-Sitte}}}, \bibinfo {author}
  {\bibfnamefont {A.}~\bibnamefont {Fert}}, \bibinfo {author} {\bibfnamefont
  {M.}~\bibnamefont {Garst}}, \bibinfo {author} {\bibfnamefont
  {T.}~\bibnamefont {Ma}}, \bibinfo {author} {\bibfnamefont {S.}~\bibnamefont
  {Mankovsky}}, \bibinfo {author} {\bibfnamefont {T.~L.}\ \bibnamefont
  {Monchesky}}, \bibinfo {author} {\bibfnamefont {M.}~\bibnamefont {Mostovoy}},
  \bibinfo {author} {\bibfnamefont {N.}~\bibnamefont {Nagaosa}}, \bibinfo
  {author} {\bibfnamefont {S.~S.~P.}\ \bibnamefont {Parkin}}, \bibinfo {author}
  {\bibfnamefont {C.}~\bibnamefont {Pfleiderer}}, \bibinfo {author}
  {\bibfnamefont {N.}~\bibnamefont {Reyren}}, \bibinfo {author} {\bibfnamefont
  {A.}~\bibnamefont {Rosch}}, \bibinfo {author} {\bibfnamefont
  {Y.}~\bibnamefont {Taguchi}},\ and\ \bibinfo {author} {\bibfnamefont
  {Y.}~\bibnamefont {Tokura}},\ }\bibfield  {title} {\bibinfo {title} {The 2020
  skyrmionics roadmap},\ }\href@noop {} {\bibfield  {journal} {\bibinfo
  {journal} {Appl. Phys.}\ ,\ \bibinfo {pages} {38}} (\bibinfo {year}
  {2020})}\BibitemShut {NoStop}%
\bibitem [{\citenamefont {Skyrme}(1961)}]{Skyrme1961a}%
  \BibitemOpen
  \bibfield  {author} {\bibinfo {author} {\bibfnamefont {T.~H.~R.}\
  \bibnamefont {Skyrme}},\ }\bibfield  {title} {\bibinfo {title} {A
  {{Non-Linear Field Theory}}},\ }\href@noop {} {\bibfield  {journal} {\bibinfo
   {journal} {Proceedings of the Royal Society of London. Series A,
  Mathematical and Physical Sciences}\ }\textbf {\bibinfo {volume} {260}},\
  \bibinfo {pages} {127} (\bibinfo {year} {1961})}\BibitemShut {NoStop}%
\bibitem [{\citenamefont {Skyrme}(1962)}]{Skyrme1962}%
  \BibitemOpen
  \bibfield  {author} {\bibinfo {author} {\bibfnamefont {T.~H.~R.}\
  \bibnamefont {Skyrme}},\ }\bibfield  {title} {\bibinfo {title} {A unified
  field theory of mesons and baryons},\ }\href
  {https://doi.org/10.1016/0029-5582(62)90775-7} {\bibfield  {journal}
  {\bibinfo  {journal} {Nucl. Phys.}\ }\textbf {\bibinfo {volume} {31}},\
  \bibinfo {pages} {556} (\bibinfo {year} {1962})}\BibitemShut {NoStop}%
\bibitem [{\citenamefont {Witten}(1983{\natexlab{a}})}]{Witten1983}%
  \BibitemOpen
  \bibfield  {author} {\bibinfo {author} {\bibfnamefont {E.}~\bibnamefont
  {Witten}},\ }\bibfield  {title} {\bibinfo {title} {Global aspects of current
  algebra},\ }\href {https://doi.org/10.1016/0550-3213(83)90063-9} {\bibfield
  {journal} {\bibinfo  {journal} {Nucl. Phys. B}\ }\textbf {\bibinfo {volume}
  {223}},\ \bibinfo {pages} {422} (\bibinfo {year}
  {1983}{\natexlab{a}})}\BibitemShut {NoStop}%
\bibitem [{\citenamefont {Witten}(1983{\natexlab{b}})}]{Witten1983a}%
  \BibitemOpen
  \bibfield  {author} {\bibinfo {author} {\bibfnamefont {E.}~\bibnamefont
  {Witten}},\ }\bibfield  {title} {\bibinfo {title} {Current algebra, baryons,
  and quark confinement},\ }\href
  {https://doi.org/10.1016/0550-3213(83)90064-0} {\bibfield  {journal}
  {\bibinfo  {journal} {Nucl. Phys. B}\ }\textbf {\bibinfo {volume} {223}},\
  \bibinfo {pages} {433} (\bibinfo {year} {1983}{\natexlab{b}})}\BibitemShut
  {NoStop}%
\bibitem [{\citenamefont {Nitta}(2013{\natexlab{a}})}]{Nitta2013}%
  \BibitemOpen
  \bibfield  {author} {\bibinfo {author} {\bibfnamefont {M.}~\bibnamefont
  {Nitta}},\ }\bibfield  {title} {\bibinfo {title} {Matryoshka {{Skyrmions}}},\
  }\href {https://doi.org/10.1016/j.nuclphysb.2013.03.003} {\bibfield
  {journal} {\bibinfo  {journal} {Nucl. Phys. B}\ }\textbf {\bibinfo {volume}
  {872}},\ \bibinfo {pages} {62} (\bibinfo {year}
  {2013}{\natexlab{a}})}\BibitemShut {NoStop}%
\bibitem [{\citenamefont {{\v S}mejkal}\ \emph {et~al.}(2018)\citenamefont {{\v
  S}mejkal}, \citenamefont {Mokrousov}, \citenamefont {Yan},\ and\
  \citenamefont {MacDonald}}]{Smejkal2018}%
  \BibitemOpen
  \bibfield  {author} {\bibinfo {author} {\bibfnamefont {L.}~\bibnamefont {{\v
  S}mejkal}}, \bibinfo {author} {\bibfnamefont {Y.}~\bibnamefont {Mokrousov}},
  \bibinfo {author} {\bibfnamefont {B.}~\bibnamefont {Yan}},\ and\ \bibinfo
  {author} {\bibfnamefont {A.~H.}\ \bibnamefont {MacDonald}},\ }\bibfield
  {title} {\bibinfo {title} {Topological antiferromagnetic spintronics},\
  }\href {https://doi.org/10.1038/s41567-018-0064-5} {\bibfield  {journal}
  {\bibinfo  {journal} {Nat. Phys.}\ }\textbf {\bibinfo {volume} {14}},\
  \bibinfo {pages} {242} (\bibinfo {year} {2018})}\BibitemShut {NoStop}%
\bibitem [{\citenamefont {Kravchuk}\ \emph {et~al.}(2019)\citenamefont
  {Kravchuk}, \citenamefont {Gomonay}, \citenamefont {Sheka}, \citenamefont
  {Rodrigues}, \citenamefont {{Everschor-Sitte}}, \citenamefont {Sinova},
  \citenamefont {{van den Brink}},\ and\ \citenamefont
  {Gaididei}}]{Kravchuk2019}%
  \BibitemOpen
  \bibfield  {author} {\bibinfo {author} {\bibfnamefont {V.~P.}\ \bibnamefont
  {Kravchuk}}, \bibinfo {author} {\bibfnamefont {O.}~\bibnamefont {Gomonay}},
  \bibinfo {author} {\bibfnamefont {D.~D.}\ \bibnamefont {Sheka}}, \bibinfo
  {author} {\bibfnamefont {D.~R.}\ \bibnamefont {Rodrigues}}, \bibinfo {author}
  {\bibfnamefont {K.}~\bibnamefont {{Everschor-Sitte}}}, \bibinfo {author}
  {\bibfnamefont {J.}~\bibnamefont {Sinova}}, \bibinfo {author} {\bibfnamefont
  {J.}~\bibnamefont {{van den Brink}}},\ and\ \bibinfo {author} {\bibfnamefont
  {Y.}~\bibnamefont {Gaididei}},\ }\bibfield  {title} {\bibinfo {title} {Spin
  eigenexcitations of an antiferromagnetic skyrmion},\ }\href
  {https://doi.org/10.1103/PhysRevB.99.184429} {\bibfield  {journal} {\bibinfo
  {journal} {Phys. Rev. B}\ }\textbf {\bibinfo {volume} {99}},\ \bibinfo
  {pages} {184429} (\bibinfo {year} {2019})}\BibitemShut {NoStop}%
\bibitem [{\citenamefont {Hongo}\ \emph {et~al.}(2020)\citenamefont {Hongo},
  \citenamefont {Fujimori}, \citenamefont {Misumi}, \citenamefont {Nitta},\
  and\ \citenamefont {Sakai}}]{Hongo2020}%
  \BibitemOpen
  \bibfield  {author} {\bibinfo {author} {\bibfnamefont {M.}~\bibnamefont
  {Hongo}}, \bibinfo {author} {\bibfnamefont {T.}~\bibnamefont {Fujimori}},
  \bibinfo {author} {\bibfnamefont {T.}~\bibnamefont {Misumi}}, \bibinfo
  {author} {\bibfnamefont {M.}~\bibnamefont {Nitta}},\ and\ \bibinfo {author}
  {\bibfnamefont {N.}~\bibnamefont {Sakai}},\ }\bibfield  {title} {\bibinfo
  {title} {Instantons in chiral magnets},\ }\href
  {https://doi.org/10.1103/PhysRevB.101.104417} {\bibfield  {journal} {\bibinfo
   {journal} {Phys. Rev. B}\ }\textbf {\bibinfo {volume} {101}},\ \bibinfo
  {pages} {104417} (\bibinfo {year} {2020})}\BibitemShut {NoStop}%
\bibitem [{\citenamefont {G{\"o}bel}\ \emph {et~al.}(2021)\citenamefont
  {G{\"o}bel}, \citenamefont {Mertig},\ and\ \citenamefont
  {Tretiakov}}]{Gobel2021}%
  \BibitemOpen
  \bibfield  {author} {\bibinfo {author} {\bibfnamefont {B.}~\bibnamefont
  {G{\"o}bel}}, \bibinfo {author} {\bibfnamefont {I.}~\bibnamefont {Mertig}},\
  and\ \bibinfo {author} {\bibfnamefont {O.~A.}\ \bibnamefont {Tretiakov}},\
  }\bibfield  {title} {\bibinfo {title} {Beyond skyrmions: {{Review}} and
  perspectives of alternative magnetic quasiparticles},\ }\href
  {https://doi.org/10.1016/j.physrep.2020.10.001} {\bibfield  {journal}
  {\bibinfo  {journal} {Phys. Rep.}\ }\textbf {\bibinfo {volume} {895}},\
  \bibinfo {pages} {1} (\bibinfo {year} {2021})}\BibitemShut {NoStop}%
\bibitem [{\citenamefont {Nitta}(2013{\natexlab{b}})}]{Nitta2013a}%
  \BibitemOpen
  \bibfield  {author} {\bibinfo {author} {\bibfnamefont {M.}~\bibnamefont
  {Nitta}},\ }\bibfield  {title} {\bibinfo {title} {Correspondence between
  {{Skyrmions}} in 2 + 1 and 3 + 1 dimensions},\ }\href
  {https://doi.org/10.1103/PhysRevD.87.025013} {\bibfield  {journal} {\bibinfo
  {journal} {Phys. Rev. D}\ }\textbf {\bibinfo {volume} {87}},\ \bibinfo
  {pages} {025013} (\bibinfo {year} {2013}{\natexlab{b}})}\BibitemShut
  {NoStop}%
\bibitem [{\citenamefont {Xia}\ \emph {et~al.}(2017)\citenamefont {Xia},
  \citenamefont {Jin}, \citenamefont {Song}, \citenamefont {Wang},
  \citenamefont {Wang},\ and\ \citenamefont {Liu}}]{Xia2017}%
  \BibitemOpen
  \bibfield  {author} {\bibinfo {author} {\bibfnamefont {H.}~\bibnamefont
  {Xia}}, \bibinfo {author} {\bibfnamefont {C.}~\bibnamefont {Jin}}, \bibinfo
  {author} {\bibfnamefont {C.}~\bibnamefont {Song}}, \bibinfo {author}
  {\bibfnamefont {J.}~\bibnamefont {Wang}}, \bibinfo {author} {\bibfnamefont
  {J.}~\bibnamefont {Wang}},\ and\ \bibinfo {author} {\bibfnamefont
  {Q.}~\bibnamefont {Liu}},\ }\bibfield  {title} {\bibinfo {title} {Control and
  manipulation of antiferromagnetic skyrmions in racetrack},\ }\href
  {https://doi.org/10.1088/1361-6463/aa95f2} {\bibfield  {journal} {\bibinfo
  {journal} {J. Phys. D: Appl. Phys.}\ }\textbf {\bibinfo {volume} {50}},\
  \bibinfo {pages} {505005} (\bibinfo {year} {2017})}\BibitemShut {NoStop}%
\bibitem [{\citenamefont {Hoffmann}\ \emph {et~al.}(2021)\citenamefont
  {Hoffmann}, \citenamefont {M{\"u}ller}, \citenamefont {Melcher},\ and\
  \citenamefont {Bl{\"u}gel}}]{Hoffmann2021a}%
  \BibitemOpen
  \bibfield  {author} {\bibinfo {author} {\bibfnamefont {M.}~\bibnamefont
  {Hoffmann}}, \bibinfo {author} {\bibfnamefont {G.~P.}\ \bibnamefont
  {M{\"u}ller}}, \bibinfo {author} {\bibfnamefont {C.}~\bibnamefont
  {Melcher}},\ and\ \bibinfo {author} {\bibfnamefont {S.}~\bibnamefont
  {Bl{\"u}gel}},\ }\bibfield  {title} {\bibinfo {title}
  {Skyrmion-{{Antiskyrmion Racetrack Memory}} in {{Rank-One DMI Materials}}},\
  }\href {https://doi.org/10.3389/fphy.2021.769873} {\bibfield  {journal}
  {\bibinfo  {journal} {Front. Phys.}\ }\textbf {\bibinfo {volume} {9}},\
  \bibinfo {pages} {769873} (\bibinfo {year} {2021})}\BibitemShut {NoStop}%
\bibitem [{\citenamefont {Parkin}\ \emph {et~al.}(2008)\citenamefont {Parkin},
  \citenamefont {Hayashi},\ and\ \citenamefont {Thomas}}]{Parkin2008}%
  \BibitemOpen
  \bibfield  {author} {\bibinfo {author} {\bibfnamefont {S.~S.~P.}\
  \bibnamefont {Parkin}}, \bibinfo {author} {\bibfnamefont {M.}~\bibnamefont
  {Hayashi}},\ and\ \bibinfo {author} {\bibfnamefont {L.}~\bibnamefont
  {Thomas}},\ }\bibfield  {title} {\bibinfo {title} {Magnetic {{Domain-Wall
  Racetrack Memory}}},\ }\href {https://doi.org/10.1126/science.1145799}
  {\bibfield  {journal} {\bibinfo  {journal} {Science}\ }\textbf {\bibinfo
  {volume} {320}},\ \bibinfo {pages} {190} (\bibinfo {year}
  {2008})}\BibitemShut {NoStop}%
\bibitem [{\citenamefont {Yu}\ \emph {et~al.}(2021)\citenamefont {Yu},
  \citenamefont {Xiao},\ and\ \citenamefont {Schultheiss}}]{Yu2021b}%
  \BibitemOpen
  \bibfield  {author} {\bibinfo {author} {\bibfnamefont {H.}~\bibnamefont
  {Yu}}, \bibinfo {author} {\bibfnamefont {J.}~\bibnamefont {Xiao}},\ and\
  \bibinfo {author} {\bibfnamefont {H.}~\bibnamefont {Schultheiss}},\
  }\bibfield  {title} {\bibinfo {title} {Magnetic texture based magnonics},\
  }\href {https://doi.org/10.1016/j.physrep.2020.12.004} {\bibfield  {journal}
  {\bibinfo  {journal} {Phys. Rep.}\ }\textbf {\bibinfo {volume} {905}},\
  \bibinfo {pages} {1} (\bibinfo {year} {2021})}\BibitemShut {NoStop}%
\bibitem [{\citenamefont {Sch{\"u}tte}\ and\ \citenamefont
  {Garst}(2014)}]{Schutte2014a}%
  \BibitemOpen
  \bibfield  {author} {\bibinfo {author} {\bibfnamefont {C.}~\bibnamefont
  {Sch{\"u}tte}}\ and\ \bibinfo {author} {\bibfnamefont {M.}~\bibnamefont
  {Garst}},\ }\bibfield  {title} {\bibinfo {title} {Magnon-skyrmion scattering
  in chiral magnets},\ }\href {https://doi.org/10.1103/PhysRevB.90.094423}
  {\bibfield  {journal} {\bibinfo  {journal} {Phys. Rev. B}\ }\textbf {\bibinfo
  {volume} {90}},\ \bibinfo {pages} {094423} (\bibinfo {year}
  {2014})}\BibitemShut {NoStop}%
\bibitem [{\citenamefont {Schroeter}\ and\ \citenamefont
  {Garst}(2015)}]{Schroeter2015}%
  \BibitemOpen
  \bibfield  {author} {\bibinfo {author} {\bibfnamefont {S.}~\bibnamefont
  {Schroeter}}\ and\ \bibinfo {author} {\bibfnamefont {M.}~\bibnamefont
  {Garst}},\ }\bibfield  {title} {\bibinfo {title} {Scattering of high-energy
  magnons off a magnetic skyrmion},\ }\href {https://doi.org/10.1063/1.4932356}
  {\bibfield  {journal} {\bibinfo  {journal} {Low Temp. Phys.}\ }\textbf
  {\bibinfo {volume} {41}},\ \bibinfo {pages} {817} (\bibinfo {year}
  {2015})}\BibitemShut {NoStop}%
\bibitem [{\citenamefont {Guslienko}(2016)}]{Guslienko2016}%
  \BibitemOpen
  \bibfield  {author} {\bibinfo {author} {\bibfnamefont {K.~Y.}\ \bibnamefont
  {Guslienko}},\ }\bibfield  {title} {\bibinfo {title} {Gauge and emergent
  electromagnetic fields for moving magnetic topological solitons},\ }\href
  {https://doi.org/10.1209/0295-5075/113/67002} {\bibfield  {journal} {\bibinfo
   {journal} {EPL}\ }\textbf {\bibinfo {volume} {113}},\ \bibinfo {pages}
  {67002} (\bibinfo {year} {2016})}\BibitemShut {NoStop}%
\bibitem [{\citenamefont {Tatara}(2019)}]{Tatara2019a}%
  \BibitemOpen
  \bibfield  {author} {\bibinfo {author} {\bibfnamefont {G.}~\bibnamefont
  {Tatara}},\ }\bibfield  {title} {\bibinfo {title} {Effective gauge field
  theory of spintronics},\ }\href {https://doi.org/10.1016/j.physe.2018.05.011}
  {\bibfield  {journal} {\bibinfo  {journal} {Physica E}\ }\textbf {\bibinfo
  {volume} {106}},\ \bibinfo {pages} {208} (\bibinfo {year}
  {2019})}\BibitemShut {NoStop}%
\bibitem [{\citenamefont {Cheng}\ and\ \citenamefont {Niu}(2012)}]{Cheng2012}%
  \BibitemOpen
  \bibfield  {author} {\bibinfo {author} {\bibfnamefont {R.}~\bibnamefont
  {Cheng}}\ and\ \bibinfo {author} {\bibfnamefont {Q.}~\bibnamefont {Niu}},\
  }\bibfield  {title} {\bibinfo {title} {Electron dynamics in slowly varying
  antiferromagnetic texture},\ }\href
  {https://doi.org/10.1103/PhysRevB.86.245118} {\bibfield  {journal} {\bibinfo
  {journal} {Phys. Rev. B}\ }\textbf {\bibinfo {volume} {86}},\ \bibinfo
  {pages} {245118} (\bibinfo {year} {2012})}\BibitemShut {NoStop}%
\bibitem [{\citenamefont {{van Hoogdalem}}\ \emph {et~al.}(2013)\citenamefont
  {{van Hoogdalem}}, \citenamefont {Tserkovnyak},\ and\ \citenamefont
  {Loss}}]{vanHoogdalem2013}%
  \BibitemOpen
  \bibfield  {author} {\bibinfo {author} {\bibfnamefont {K.~A.}\ \bibnamefont
  {{van Hoogdalem}}}, \bibinfo {author} {\bibfnamefont {Y.}~\bibnamefont
  {Tserkovnyak}},\ and\ \bibinfo {author} {\bibfnamefont {D.}~\bibnamefont
  {Loss}},\ }\bibfield  {title} {\bibinfo {title} {Magnetic texture-induced
  thermal {{Hall}} effects},\ }\href
  {https://doi.org/10.1103/PhysRevB.87.024402} {\bibfield  {journal} {\bibinfo
  {journal} {Phys. Rev. B}\ }\textbf {\bibinfo {volume} {87}},\ \bibinfo
  {pages} {024402} (\bibinfo {year} {2013})}\BibitemShut {NoStop}%
\bibitem [{\citenamefont {Li}\ and\ \citenamefont {Kovalev}(2021)}]{Li2021b}%
  \BibitemOpen
  \bibfield  {author} {\bibinfo {author} {\bibfnamefont {B.}~\bibnamefont
  {Li}}\ and\ \bibinfo {author} {\bibfnamefont {A.~A.}\ \bibnamefont
  {Kovalev}},\ }\bibfield  {title} {\bibinfo {title} {Spin superfluidity in
  noncollinear antiferromagnets},\ }\href
  {https://doi.org/10.1103/PhysRevB.103.L060406} {\bibfield  {journal}
  {\bibinfo  {journal} {Phys. Rev. B}\ }\textbf {\bibinfo {volume} {103}},\
  \bibinfo {pages} {L060406} (\bibinfo {year} {2021})}\BibitemShut {NoStop}%
\bibitem [{\citenamefont {Lee}\ and\ \citenamefont {Kim}(2021)}]{Lee2021}%
  \BibitemOpen
  \bibfield  {author} {\bibinfo {author} {\bibfnamefont {S.}~\bibnamefont
  {Lee}}\ and\ \bibinfo {author} {\bibfnamefont {S.~K.}\ \bibnamefont {Kim}},\
  }\bibfield  {title} {\bibinfo {title} {Orbital angular momentum and
  current-induced motion of a topologically textured domain wall in a
  ferromagnetic nanotube},\ }\href
  {https://doi.org/10.1103/PhysRevB.104.L140401} {\bibfield  {journal}
  {\bibinfo  {journal} {Phys. Rev. B}\ }\textbf {\bibinfo {volume} {104}},\
  \bibinfo {pages} {L140401} (\bibinfo {year} {2021})}\BibitemShut {NoStop}%
\bibitem [{\citenamefont {Lee}\ and\ \citenamefont {Kim}(2022)}]{Lee2022}%
  \BibitemOpen
  \bibfield  {author} {\bibinfo {author} {\bibfnamefont {S.}~\bibnamefont
  {Lee}}\ and\ \bibinfo {author} {\bibfnamefont {S.~K.}\ \bibnamefont {Kim}},\
  }\bibfield  {title} {\bibinfo {title} {Generation of {{Magnon Orbital Angular
  Momentum}} by a {{Skyrmion-Textured Domain Wall}} in a {{Ferromagnetic
  Nanotube}}},\ }\href {https://doi.org/10.3389/fphy.2022.858614} {\bibfield
  {journal} {\bibinfo  {journal} {Front. Phys.}\ }\textbf {\bibinfo {volume}
  {10}},\ \bibinfo {pages} {858614} (\bibinfo {year} {2022})}\BibitemShut
  {NoStop}%
\bibitem [{\citenamefont {Yan}\ \emph {et~al.}(2011)\citenamefont {Yan},
  \citenamefont {Wang},\ and\ \citenamefont {Wang}}]{Yan2011a}%
  \BibitemOpen
  \bibfield  {author} {\bibinfo {author} {\bibfnamefont {P.}~\bibnamefont
  {Yan}}, \bibinfo {author} {\bibfnamefont {X.~S.}\ \bibnamefont {Wang}},\ and\
  \bibinfo {author} {\bibfnamefont {X.~R.}\ \bibnamefont {Wang}},\ }\bibfield
  {title} {\bibinfo {title} {All-{{Magnonic Spin-Transfer Torque}} and {{Domain
  Wall Propagation}}},\ }\href {https://doi.org/10.1103/PhysRevLett.107.177207}
  {\bibfield  {journal} {\bibinfo  {journal} {Phys. Rev. Lett.}\ }\textbf
  {\bibinfo {volume} {107}},\ \bibinfo {pages} {177207} (\bibinfo {year}
  {2011})}\BibitemShut {NoStop}%
\bibitem [{\citenamefont {Kim}\ \emph {et~al.}(2014)\citenamefont {Kim},
  \citenamefont {Tserkovnyak},\ and\ \citenamefont {Tchernyshyov}}]{Kim2014}%
  \BibitemOpen
  \bibfield  {author} {\bibinfo {author} {\bibfnamefont {S.~K.}\ \bibnamefont
  {Kim}}, \bibinfo {author} {\bibfnamefont {Y.}~\bibnamefont {Tserkovnyak}},\
  and\ \bibinfo {author} {\bibfnamefont {O.}~\bibnamefont {Tchernyshyov}},\
  }\bibfield  {title} {\bibinfo {title} {Propulsion of a domain wall in an
  antiferromagnet by magnons},\ }\href
  {https://doi.org/10.1103/PhysRevB.90.104406} {\bibfield  {journal} {\bibinfo
  {journal} {Phys. Rev. B}\ }\textbf {\bibinfo {volume} {90}},\ \bibinfo
  {pages} {104406} (\bibinfo {year} {2014})}\BibitemShut {NoStop}%
\bibitem [{\citenamefont {Sukumar}(1985)}]{Sukumar1985}%
  \BibitemOpen
  \bibfield  {author} {\bibinfo {author} {\bibfnamefont {C.~V.}\ \bibnamefont
  {Sukumar}},\ }\bibfield  {title} {\bibinfo {title} {Supersymmetric quantum
  mechanics of one-dimensional systems},\ }\href
  {https://doi.org/10.1088/0305-4470/18/15/020} {\bibfield  {journal} {\bibinfo
   {journal} {J. Phys. A: Math. Gen.}\ }\textbf {\bibinfo {volume} {18}},\
  \bibinfo {pages} {2917} (\bibinfo {year} {1985})}\BibitemShut {NoStop}%
\bibitem [{\citenamefont {Cooper}\ \emph {et~al.}(1995)\citenamefont {Cooper},
  \citenamefont {Khare},\ and\ \citenamefont {Sukhatme}}]{Cooper1995}%
  \BibitemOpen
  \bibfield  {author} {\bibinfo {author} {\bibfnamefont {F.}~\bibnamefont
  {Cooper}}, \bibinfo {author} {\bibfnamefont {A.}~\bibnamefont {Khare}},\ and\
  \bibinfo {author} {\bibfnamefont {U.}~\bibnamefont {Sukhatme}},\ }\bibfield
  {title} {\bibinfo {title} {Supersymmetry and quantum mechanics},\ }\href
  {https://doi.org/10.1016/0370-1573(94)00080-M} {\bibfield  {journal}
  {\bibinfo  {journal} {Phys. Rep.}\ }\textbf {\bibinfo {volume} {251}},\
  \bibinfo {pages} {267} (\bibinfo {year} {1995})}\BibitemShut {NoStop}%
\bibitem [{\citenamefont {Go}\ \emph {et~al.}(2022)\citenamefont {Go},
  \citenamefont {Lee},\ and\ \citenamefont {Kim}}]{Go2022}%
  \BibitemOpen
  \bibfield  {author} {\bibinfo {author} {\bibfnamefont {G.}~\bibnamefont
  {Go}}, \bibinfo {author} {\bibfnamefont {S.}~\bibnamefont {Lee}},\ and\
  \bibinfo {author} {\bibfnamefont {S.~K.}\ \bibnamefont {Kim}},\ }\bibfield
  {title} {\bibinfo {title} {Generation of nonreciprocity in gapless spin waves
  by chirality injection},\ }\href
  {https://doi.org/10.1103/PhysRevB.105.134401} {\bibfield  {journal} {\bibinfo
   {journal} {Phys. Rev. B}\ }\textbf {\bibinfo {volume} {105}},\ \bibinfo
  {pages} {134401} (\bibinfo {year} {2022})}\BibitemShut {NoStop}%
\bibitem [{\citenamefont {Tchrakian}\ and\ \citenamefont
  {Arthur}(1995)}]{Tchrakian1995}%
  \BibitemOpen
  \bibfield  {author} {\bibinfo {author} {\bibfnamefont {D.}~\bibnamefont
  {Tchrakian}}\ and\ \bibinfo {author} {\bibfnamefont {K.}~\bibnamefont
  {Arthur}},\ }\bibfield  {title} {\bibinfo {title} {Solitons in gauged
  {{Sigma}} models: 2 dimensions},\ }\href
  {https://doi.org/10.1016/0370-2693(95)00448-T} {\bibfield  {journal}
  {\bibinfo  {journal} {Phys. Lett. B}\ }\textbf {\bibinfo {volume} {352}},\
  \bibinfo {pages} {327} (\bibinfo {year} {1995})}\BibitemShut {NoStop}%
\bibitem [{\citenamefont {Schroers}(2019)}]{Schroers2019}%
  \BibitemOpen
  \bibfield  {author} {\bibinfo {author} {\bibfnamefont {B.}~\bibnamefont
  {Schroers}},\ }\bibfield  {title} {\bibinfo {title} {Gauged sigma models and
  magnetic {{Skyrmions}}},\ }\href
  {https://doi.org/10.21468/SciPostPhys.7.3.030} {\bibfield  {journal}
  {\bibinfo  {journal} {SciPost Phys.}\ }\textbf {\bibinfo {volume} {7}},\
  \bibinfo {pages} {030} (\bibinfo {year} {2019})}\BibitemShut {NoStop}%
\bibitem [{\citenamefont {Speight}\ and\ \citenamefont
  {Winyard}(2020)}]{Speight2020}%
  \BibitemOpen
  \bibfield  {author} {\bibinfo {author} {\bibfnamefont {M.}~\bibnamefont
  {Speight}}\ and\ \bibinfo {author} {\bibfnamefont {T.}~\bibnamefont
  {Winyard}},\ }\bibfield  {title} {\bibinfo {title} {Skyrmions and spin waves
  in frustrated ferromagnets at low applied magnetic field},\ }\href
  {https://doi.org/10.1103/PhysRevB.101.134420} {\bibfield  {journal} {\bibinfo
   {journal} {Phys. Rev. B}\ }\textbf {\bibinfo {volume} {101}},\ \bibinfo
  {pages} {134420} (\bibinfo {year} {2020})}\BibitemShut {NoStop}%
\bibitem [{\citenamefont {Ivanov}(1983)}]{Ivanov1983}%
  \BibitemOpen
  \bibfield  {author} {\bibinfo {author} {\bibfnamefont {G.~G.}\ \bibnamefont
  {Ivanov}},\ }\bibfield  {title} {\bibinfo {title} {Symmetries, conservation
  laws, and exact solutions in nonlinear sigma models},\ }\href
  {https://doi.org/10.1007/BF01028173} {\bibfield  {journal} {\bibinfo
  {journal} {Theor. Math. Phys.}\ }\textbf {\bibinfo {volume} {57}},\ \bibinfo
  {pages} {981} (\bibinfo {year} {1983})}\BibitemShut {NoStop}%
\bibitem [{\citenamefont {Haldane}(1983{\natexlab{a}})}]{Haldane1983}%
  \BibitemOpen
  \bibfield  {author} {\bibinfo {author} {\bibfnamefont {F.~D.~M.}\
  \bibnamefont {Haldane}},\ }\bibfield  {title} {\bibinfo {title} {Nonlinear
  {{Field Theory}} of {{Large-Spin Heisenberg Antiferromagnets}}:
  {{Semiclassically Quantized Solitons}} of the {{One-Dimensional Easy-Axis
  N\'eel State}}},\ }\href {https://doi.org/10.1103/PhysRevLett.50.1153}
  {\bibfield  {journal} {\bibinfo  {journal} {Phys. Rev. Lett.}\ }\textbf
  {\bibinfo {volume} {50}},\ \bibinfo {pages} {1153} (\bibinfo {year}
  {1983}{\natexlab{a}})}\BibitemShut {NoStop}%
\bibitem [{\citenamefont {Haldane}(1983{\natexlab{b}})}]{Haldane1983a}%
  \BibitemOpen
  \bibfield  {author} {\bibinfo {author} {\bibfnamefont {F.~D.~M.}\
  \bibnamefont {Haldane}},\ }\bibfield  {title} {\bibinfo {title} {Continuum
  dynamics of the 1-{{D Heisenberg}} antiferromagnet: {{Identification}} with
  the {{O}}(3) nonlinear sigma model},\ }\href
  {https://doi.org/10.1016/0375-9601(83)90631-X} {\bibfield  {journal}
  {\bibinfo  {journal} {Phys. lett., A}\ }\textbf {\bibinfo {volume} {93}},\
  \bibinfo {pages} {464} (\bibinfo {year} {1983}{\natexlab{b}})}\BibitemShut
  {NoStop}%
\bibitem [{\citenamefont {Hasselmann}\ and\ \citenamefont
  {Kopietz}(2006)}]{Hasselmann2006}%
  \BibitemOpen
  \bibfield  {author} {\bibinfo {author} {\bibfnamefont {N.}~\bibnamefont
  {Hasselmann}}\ and\ \bibinfo {author} {\bibfnamefont {P.}~\bibnamefont
  {Kopietz}},\ }\bibfield  {title} {\bibinfo {title} {Spin-wave interactions in
  quantum antiferromagnets},\ }\href
  {https://doi.org/10.1209/epl/i2006-10060-6} {\bibfield  {journal} {\bibinfo
  {journal} {EPL}\ }\textbf {\bibinfo {volume} {74}},\ \bibinfo {pages} {1067}
  (\bibinfo {year} {2006})}\BibitemShut {NoStop}%
\bibitem [{\citenamefont {Kobayashi}\ and\ \citenamefont
  {Nitta}(2013)}]{Kobayashi2013a}%
  \BibitemOpen
  \bibfield  {author} {\bibinfo {author} {\bibfnamefont {M.}~\bibnamefont
  {Kobayashi}}\ and\ \bibinfo {author} {\bibfnamefont {M.}~\bibnamefont
  {Nitta}},\ }\bibfield  {title} {\bibinfo {title} {Sine-{{Gordon}} kinks on a
  domain wall ring},\ }\href {https://doi.org/10.1103/PhysRevD.87.085003}
  {\bibfield  {journal} {\bibinfo  {journal} {Phys. Rev. D}\ }\textbf {\bibinfo
  {volume} {87}},\ \bibinfo {pages} {085003} (\bibinfo {year}
  {2013})}\BibitemShut {NoStop}%
\bibitem [{\citenamefont {Kim}\ and\ \citenamefont
  {Tserkovnyak}(2017{\natexlab{a}})}]{Kim2017a}%
  \BibitemOpen
  \bibfield  {author} {\bibinfo {author} {\bibfnamefont {S.~K.}\ \bibnamefont
  {Kim}}\ and\ \bibinfo {author} {\bibfnamefont {Y.}~\bibnamefont
  {Tserkovnyak}},\ }\bibfield  {title} {\bibinfo {title} {Chiral {{Edge Mode}}
  in the {{Coupled Dynamics}} of {{Magnetic Solitons}} in a {{Honeycomb
  Lattice}}},\ }\href {https://doi.org/10.1103/PhysRevLett.119.077204}
  {\bibfield  {journal} {\bibinfo  {journal} {Phys. Rev. Lett.}\ }\textbf
  {\bibinfo {volume} {119}},\ \bibinfo {pages} {077204} (\bibinfo {year}
  {2017}{\natexlab{a}})}\BibitemShut {NoStop}%
\bibitem [{\citenamefont {Flebus}\ \emph {et~al.}(2018)\citenamefont {Flebus},
  \citenamefont {Ochoa}, \citenamefont {Upadhyaya},\ and\ \citenamefont
  {Tserkovnyak}}]{Flebus2018}%
  \BibitemOpen
  \bibfield  {author} {\bibinfo {author} {\bibfnamefont {B.}~\bibnamefont
  {Flebus}}, \bibinfo {author} {\bibfnamefont {H.}~\bibnamefont {Ochoa}},
  \bibinfo {author} {\bibfnamefont {P.}~\bibnamefont {Upadhyaya}},\ and\
  \bibinfo {author} {\bibfnamefont {Y.}~\bibnamefont {Tserkovnyak}},\
  }\bibfield  {title} {\bibinfo {title} {Proposal for dynamic imaging of
  antiferromagnetic domain wall via quantum-impurity relaxometry},\ }\href
  {https://doi.org/10.1103/PhysRevB.98.180409} {\bibfield  {journal} {\bibinfo
  {journal} {Phys. Rev. B}\ }\textbf {\bibinfo {volume} {98}},\ \bibinfo
  {pages} {180409} (\bibinfo {year} {2018})}\BibitemShut {NoStop}%
\bibitem [{\citenamefont {Flebus}\ and\ \citenamefont
  {Tserkovnyak}(2019)}]{Flebus2019}%
  \BibitemOpen
  \bibfield  {author} {\bibinfo {author} {\bibfnamefont {B.}~\bibnamefont
  {Flebus}}\ and\ \bibinfo {author} {\bibfnamefont {Y.}~\bibnamefont
  {Tserkovnyak}},\ }\bibfield  {title} {\bibinfo {title} {Entangling distant
  spin qubits via a magnetic domain wall},\ }\href
  {https://doi.org/10.1103/PhysRevB.99.140403} {\bibfield  {journal} {\bibinfo
  {journal} {Phys. Rev. B}\ }\textbf {\bibinfo {volume} {99}},\ \bibinfo
  {pages} {140403} (\bibinfo {year} {2019})}\BibitemShut {NoStop}%
\bibitem [{\citenamefont {Li}\ and\ \citenamefont {Kovalev}(2020)}]{Li2020}%
  \BibitemOpen
  \bibfield  {author} {\bibinfo {author} {\bibfnamefont {B.}~\bibnamefont
  {Li}}\ and\ \bibinfo {author} {\bibfnamefont {A.~A.}\ \bibnamefont
  {Kovalev}},\ }\bibfield  {title} {\bibinfo {title} {Magnon {{Landau Levels}}
  and {{Spin Responses}} in {{Antiferromagnets}}},\ }\href
  {https://doi.org/10.1103/PhysRevLett.125.257201} {\bibfield  {journal}
  {\bibinfo  {journal} {Phys. Rev. Lett.}\ }\textbf {\bibinfo {volume} {125}},\
  \bibinfo {pages} {257201} (\bibinfo {year} {2020})}\BibitemShut {NoStop}%
\bibitem [{\citenamefont {Rosen}\ and\ \citenamefont
  {Morse}(1932)}]{Rosen1932}%
  \BibitemOpen
  \bibfield  {author} {\bibinfo {author} {\bibfnamefont {N.}~\bibnamefont
  {Rosen}}\ and\ \bibinfo {author} {\bibfnamefont {P.~M.}\ \bibnamefont
  {Morse}},\ }\bibfield  {title} {\bibinfo {title} {On the {{Vibrations}} of
  {{Polyatomic Molecules}}},\ }\href {https://doi.org/10.1103/PhysRev.42.210}
  {\bibfield  {journal} {\bibinfo  {journal} {Phys. Rev.}\ }\textbf {\bibinfo
  {volume} {42}},\ \bibinfo {pages} {210} (\bibinfo {year} {1932})}\BibitemShut
  {NoStop}%
\bibitem [{\citenamefont {P{\"o}schl}\ and\ \citenamefont
  {Teller}(1933)}]{Poeschl1933}%
  \BibitemOpen
  \bibfield  {author} {\bibinfo {author} {\bibfnamefont {G.}~\bibnamefont
  {P{\"o}schl}}\ and\ \bibinfo {author} {\bibfnamefont {E.}~\bibnamefont
  {Teller}},\ }\bibfield  {title} {\bibinfo {title} {{Bemerkungen zur
  Quantenmechanik des anharmonischen Oszillators}},\ }\href
  {https://doi.org/10.1007/BF01331132} {\bibfield  {journal} {\bibinfo
  {journal} {Z. Physik}\ }\textbf {\bibinfo {volume} {83}},\ \bibinfo {pages}
  {143} (\bibinfo {year} {1933})}\BibitemShut {NoStop}%
\bibitem [{\citenamefont {Lekner}(2007)}]{Lekner2007}%
  \BibitemOpen
  \bibfield  {author} {\bibinfo {author} {\bibfnamefont {J.}~\bibnamefont
  {Lekner}},\ }\bibfield  {title} {\bibinfo {title} {Reflectionless eigenstates
  of the sech2 potential},\ }\href {https://doi.org/10.1119/1.2787015}
  {\bibfield  {journal} {\bibinfo  {journal} {Am. J. Phys.}\ }\textbf {\bibinfo
  {volume} {75}},\ \bibinfo {pages} {1151} (\bibinfo {year}
  {2007})}\BibitemShut {NoStop}%
\bibitem [{\citenamefont {Zhang}\ \emph {et~al.}(2018)\citenamefont {Zhang},
  \citenamefont {Wang}, \citenamefont {Cao}, \citenamefont {Yan},\ and\
  \citenamefont {Wang}}]{Zhang2018c}%
  \BibitemOpen
  \bibfield  {author} {\bibinfo {author} {\bibfnamefont {B.}~\bibnamefont
  {Zhang}}, \bibinfo {author} {\bibfnamefont {Z.}~\bibnamefont {Wang}},
  \bibinfo {author} {\bibfnamefont {Y.}~\bibnamefont {Cao}}, \bibinfo {author}
  {\bibfnamefont {P.}~\bibnamefont {Yan}},\ and\ \bibinfo {author}
  {\bibfnamefont {X.~R.}\ \bibnamefont {Wang}},\ }\bibfield  {title} {\bibinfo
  {title} {Eavesdropping on spin waves inside the domain-wall nanochannel via
  three-magnon processes},\ }\href {https://doi.org/10.1103/PhysRevB.97.094421}
  {\bibfield  {journal} {\bibinfo  {journal} {Phys. Rev. B}\ }\textbf {\bibinfo
  {volume} {97}},\ \bibinfo {pages} {094421} (\bibinfo {year}
  {2018})}\BibitemShut {NoStop}%
\bibitem [{\citenamefont {Gadella}\ \emph {et~al.}(2017)\citenamefont
  {Gadella}, \citenamefont {Kuru},\ and\ \citenamefont {Negro}}]{Gadella2017}%
  \BibitemOpen
  \bibfield  {author} {\bibinfo {author} {\bibfnamefont {M.}~\bibnamefont
  {Gadella}}, \bibinfo {author} {\bibfnamefont {{\c S}.}~\bibnamefont {Kuru}},\
  and\ \bibinfo {author} {\bibfnamefont {J.}~\bibnamefont {Negro}},\ }\bibfield
   {title} {\bibinfo {title} {The hyperbolic step potential: {{Anti-bound}}
  states, {{SUSY}} partners and {{Wigner}} time delays},\ }\href
  {https://doi.org/10.1016/j.aop.2017.02.013} {\bibfield  {journal} {\bibinfo
  {journal} {Ann. Phys.}\ }\textbf {\bibinfo {volume} {379}},\ \bibinfo {pages}
  {86} (\bibinfo {year} {2017})}\BibitemShut {NoStop}%
\bibitem [{\citenamefont {Boonserm}\ and\ \citenamefont
  {Visser}(2011)}]{Boonserm2011}%
  \BibitemOpen
  \bibfield  {author} {\bibinfo {author} {\bibfnamefont {P.}~\bibnamefont
  {Boonserm}}\ and\ \bibinfo {author} {\bibfnamefont {M.}~\bibnamefont
  {Visser}},\ }\bibfield  {title} {\bibinfo {title} {Quasi-normal frequencies:
  Key analytic results},\ }\href {https://doi.org/10.1007/JHEP03(2011)073}
  {\bibfield  {journal} {\bibinfo  {journal} {J. High Energ. Phys.}\ }\textbf
  {\bibinfo {volume} {2011}}\bibinfo  {number} { (3)},\ \bibinfo {pages}
  {73}}\BibitemShut {NoStop}%
\bibitem [{\citenamefont {Kim}\ \emph {et~al.}(2019)\citenamefont {Kim},
  \citenamefont {Nakata}, \citenamefont {Loss},\ and\ \citenamefont
  {Tserkovnyak}}]{Kim2019b}%
  \BibitemOpen
\bibfield  {number} {  }\bibfield  {author} {\bibinfo {author} {\bibfnamefont
  {S.~K.}\ \bibnamefont {Kim}}, \bibinfo {author} {\bibfnamefont
  {K.}~\bibnamefont {Nakata}}, \bibinfo {author} {\bibfnamefont
  {D.}~\bibnamefont {Loss}},\ and\ \bibinfo {author} {\bibfnamefont
  {Y.}~\bibnamefont {Tserkovnyak}},\ }\bibfield  {title} {\bibinfo {title}
  {Tunable {{Magnonic Thermal Hall Effect}} in {{Skyrmion Crystal Phases}} of
  {{Ferrimagnets}}},\ }\href {https://doi.org/10.1103/PhysRevLett.122.057204}
  {\bibfield  {journal} {\bibinfo  {journal} {Phys. Rev. Lett.}\ }\textbf
  {\bibinfo {volume} {122}},\ \bibinfo {pages} {057204} (\bibinfo {year}
  {2019})}\BibitemShut {NoStop}%
\bibitem [{\citenamefont {Kim}\ and\ \citenamefont
  {Tserkovnyak}(2017{\natexlab{b}})}]{Kim2017}%
  \BibitemOpen
  \bibfield  {author} {\bibinfo {author} {\bibfnamefont {S.~K.}\ \bibnamefont
  {Kim}}\ and\ \bibinfo {author} {\bibfnamefont {Y.}~\bibnamefont
  {Tserkovnyak}},\ }\bibfield  {title} {\bibinfo {title} {Magnetic {{Domain
  Walls}} as {{Hosts}} of {{Spin Superfluids}} and {{Generators}} of
  {{Skyrmions}}},\ }\href {https://doi.org/10.1103/PhysRevLett.119.047202}
  {\bibfield  {journal} {\bibinfo  {journal} {Phys. Rev. Lett.}\ }\textbf
  {\bibinfo {volume} {119}},\ \bibinfo {pages} {047202} (\bibinfo {year}
  {2017}{\natexlab{b}})}\BibitemShut {NoStop}%
\bibitem [{\citenamefont {Shen}\ \emph {et~al.}(2020)\citenamefont {Shen},
  \citenamefont {Tserkovnyak},\ and\ \citenamefont {Kim}}]{Shen2020a}%
  \BibitemOpen
  \bibfield  {author} {\bibinfo {author} {\bibfnamefont {P.}~\bibnamefont
  {Shen}}, \bibinfo {author} {\bibfnamefont {Y.}~\bibnamefont {Tserkovnyak}},\
  and\ \bibinfo {author} {\bibfnamefont {S.~K.}\ \bibnamefont {Kim}},\
  }\bibfield  {title} {\bibinfo {title} {Driving a magnetized domain wall in an
  antiferromagnet by magnons},\ }\href {https://doi.org/10.1063/5.0006038}
  {\bibfield  {journal} {\bibinfo  {journal} {J. Appl. Phys.}\ }\textbf
  {\bibinfo {volume} {127}},\ \bibinfo {pages} {223905} (\bibinfo {year}
  {2020})}\BibitemShut {NoStop}%
\bibitem [{\citenamefont {Yan}\ and\ \citenamefont {Bauer}(2012)}]{Yan2012}%
  \BibitemOpen
  \bibfield  {author} {\bibinfo {author} {\bibfnamefont {P.}~\bibnamefont
  {Yan}}\ and\ \bibinfo {author} {\bibfnamefont {G.~E.~W.}\ \bibnamefont
  {Bauer}},\ }\bibfield  {title} {\bibinfo {title} {Magnonic {{Domain Wall Heat
  Conductance}} in {{Ferromagnetic Wires}}},\ }\href
  {https://doi.org/10.1103/PhysRevLett.109.087202} {\bibfield  {journal}
  {\bibinfo  {journal} {Phys. Rev. Lett.}\ }\textbf {\bibinfo {volume} {109}},\
  \bibinfo {pages} {087202} (\bibinfo {year} {2012})}\BibitemShut {NoStop}%
\bibitem [{\citenamefont {Pekola}\ and\ \citenamefont
  {Karimi}(2021)}]{Pekola2021}%
  \BibitemOpen
  \bibfield  {author} {\bibinfo {author} {\bibfnamefont {J.~P.}\ \bibnamefont
  {Pekola}}\ and\ \bibinfo {author} {\bibfnamefont {B.}~\bibnamefont
  {Karimi}},\ }\bibfield  {title} {\bibinfo {title} {{\emph{Colloquium}} :
  {{Quantum}} heat transport in condensed matter systems},\ }\href
  {https://doi.org/10.1103/RevModPhys.93.041001} {\bibfield  {journal}
  {\bibinfo  {journal} {Rev. Mod. Phys.}\ }\textbf {\bibinfo {volume} {93}},\
  \bibinfo {pages} {041001} (\bibinfo {year} {2021})}\BibitemShut {NoStop}%
\bibitem [{\citenamefont {Hill}\ \emph {et~al.}(2021)\citenamefont {Hill},
  \citenamefont {Slastikov},\ and\ \citenamefont {Tchernyshyov}}]{Hill2021}%
  \BibitemOpen
  \bibfield  {author} {\bibinfo {author} {\bibfnamefont {D.}~\bibnamefont
  {Hill}}, \bibinfo {author} {\bibfnamefont {V.}~\bibnamefont {Slastikov}},\
  and\ \bibinfo {author} {\bibfnamefont {O.}~\bibnamefont {Tchernyshyov}},\
  }\bibfield  {title} {\bibinfo {title} {Chiral magnetism: A geometric
  perspective},\ }\href {https://doi.org/10.21468/SciPostPhys.10.3.078}
  {\bibfield  {journal} {\bibinfo  {journal} {SciPost Phys.}\ }\textbf
  {\bibinfo {volume} {10}},\ \bibinfo {pages} {078} (\bibinfo {year}
  {2021})}\BibitemShut {NoStop}%
\end{thebibliography}%

\end{document}